\documentclass[titlepage,12pt]{article}
\usepackage[margin=1in]{geometry}
\usepackage{latexsym}
\usepackage{amsmath}
\usepackage{amsfonts}
\usepackage{amsthm}
\usepackage{natbib}
\usepackage{setspace}
\usepackage[usenames,dvipsnames]{color}
\usepackage{enumerate}
\usepackage{latexsym}
\usepackage{amsmath,amssymb}
\usepackage{graphicx}
\usepackage[usenames,dvipsnames]{color}
\usepackage{verbatim}

\usepackage[ruled, slide, lined]{algorithm2e}
\usepackage{graphicx}
\newtheorem{lemma}{Lemma}
\newtheorem{corollary}{Corollary}

\numberwithin{equation}{section}
\numberwithin{lemma}{section}
\numberwithin{corollary}{section}

\newcommand{\bl}{\begin{flushleft}}
\newcommand{\el}{\end{flushleft}}

\newcommand{\D}{{D}}
\newcommand{\pn}{{\mathbb{P}_n}}

\newcommand{\BX}{{X}}
\newcommand{\bX}{{  X}}
\newcommand{\Bx}{{ x}}
\newcommand{\bx}{{  x}}

\newcommand{\argmin}{\mbox{argmin}}
\newcommand{\sign}{\mbox{sign}}
\newcommand{\vol}{\mbox{Vol}}

\newcommand{\supp}{\mbox{supp}}

\def\Stas{\textcolor{blue}}

\def\m{\mathcal}
\def\mb{\mathbb}

\def\act{{\rm act}}
\def\r{\right}
\def\l{\left}

\def\eps{\varepsilon}

\newtheorem{thm}{Theorem}[section]

\begin{document}

\title{\bf \Large Active Clinical Trials for Personalized Medicine}
\author{Stanislav Minsker\thanks{Visiting Assistant Professor, Mathematics Department, Duke University, Box 90320,
Durham, NC 27708, Email: sminsker@math.duke.edu. Research sponsored by NSF grants FODAVA CCF-0808847, DMS-0847388, ATD-1222567.},
Ying-Qi Zhao\thanks{Assistant Professor, Corresponding Author, Department of Biostatistics \& Medical Informatics, University of Wisconsin, Madison, WI, 53792, Email: yqzhao@biostat.wisc.edu.}, and Guang Cheng\thanks{Associate Professor, Department of Statistics, Purdue University, West Lafayette, IN 47907, Email: chengg@purdue.edu. Research Sponsored by NSF (DMS-0906497, CAREER Award DMS-1151692, DMS-1418042), Simons Fellowship in Mathematics, Office of Naval Research (ONR N00014-15-1-2331) and a grant from Indiana Clinical and Translational Sciences Institute. Guang Cheng was visiting SAMSI and on sabbatical at Princeton while this work was carried out and revised; he would like to thank SAMSI and Princeton ORFE department for their hospitality and support. We thank Dr. A. John Rush and the investigators for use of their data from the Nefazodone CBASP trial. The stimulant data used in this article were obtained from the datasets distributed by the NIDA.}}
\date{ }
\maketitle

\begin{abstract}\noindent
Individualized treatment rules (ITRs) tailor treatments according to individual patient characteristics. They can significantly improve patient care and are thus becoming increasingly popular. The data collected during randomized clinical trials are often used to estimate the optimal ITRs. However, these trials are generally expensive to run, and, moreover, they are not designed to efficiently estimate ITRs. In this paper, we propose a cost-effective estimation method from an active learning perspective. In particular, our method recruits only the ``most informative'' patients (in terms of learning the optimal ITRs) from an ongoing clinical trial. Simulation studies and real-data examples show that our active clinical trial method significantly improves on competing methods. We derive risk bounds and show that they support these observed empirical advantages.\\\\

\end{abstract}

\section{Introduction}

It is widely recognized that different patients respond differently to the same treatment. Recent advances in personalized medicine have the potential to improve treatment decisions in clinical practice by  tailoring the clinical interventions to the patient characteristics. These characteristics include demographics, medical histories, and genetic or genomic information \citep{Hamburg:pathPM2009}. It is anticipated that these new developments in personalized medicine may salvage some failed medications, which is especially important given the low overall success rate recently observed in clinical trials \citep{dimasi2010trends}.

The success of personalized medicine is contingent on the correct identification of the best treatments for each individual. One research direction is subgroup analysis, where the patients are grouped based on the estimated individual-level treatment differences \citep{cai2011analysis,foster2011subgroup}. Alternatively, vigorous research has focused on finding optimal treatment regimens, which yield the greatest benefit overall for the whole population.
Some methods involve fitting a regression model for the response, and recommending the treatment achieving the best prediction \citep{qian:itr11}. Instead, \cite{Zhao:OWL12} explore the optimal individualized therapies from a classification perspective (see also \citet{Zhang:RobustOTR2012}). All these methods are implemented using data from randomized clinical trials (RCTs).
However, traditional RCTs are primarily designed to confirm the efficacy of new treatments; they do not generate comprehensive personalized therapy rules in an {\em efficient} manner. Consequently, post-mining data from RCTs is not ideal for finding optimal treatment strategies \citep{cui2002issues, lagakos2006challenge}. RCTs generally require a large sample size to demonstrate the efficacy of a candidate treatment, and they can be expensive to run because of the need to treat and monitor a large number of subjects. In addition, RCTs waste trial resources on subjects who experience treatment effects that are relatively large. Therefore, it would be desirable to design cost-effective clinical trials for personalized medicine. Such trials would highlight individual differences in responses and take advantage of continuing advances revealed in the trial \citep{singer2005personalized}.

We propose an active learning framework for conducting clinical trials, called {\em active clinical trials}. In these trials, patients are judiciously recruited so that the optimal ITRs can be learned with fewer patients being randomized; this is a cost-effective method. Unlike traditional RCTs, we will exclude the patients for whom the benefit from one of the treatments is clearly observed, thus concentrating on those for whom the difference is less pronounced. Specifically, within the classification framework \citep{Zhao:OWL12}, we first construct ``confidence intervals" for the optimal decision boundary using the data accumulated so far based on either a frequentist or Bayesian approach. We then selectively enroll the patients whose optimal treatments are hard to determine, i.e., their benefit differences from the different treatments are ``small,'' based on the above confidence intervals. Those patients are viewed as the most informative for the purpose of learning the optimal ITRs, and thus they are recruited for randomizations at the next stage.   As will be seen in the empirical and theoretical analysis, the real-time selection of the right patients indeed improves the chance of discovering optimal ITRs with a drastically reduced sample size and cost.

Our work is related to the topic of budgeted learning \citep{madani2004budgeted, raghavan2006active, deng2007bandit} in computer science. These work aim to find the most accurate classifier subject to a fixed budget. The main difference is that the responses for the subjects are assumed to be {\em known}, and the covariate information is required. Our work in the clinical trial setting sequentially applies treatments to selected subjects and then observes their responses. Another related topic in computer science is the ``multi-armed bandit problem'' \citep{robbins1952some}, where resources are allocated among multiple arms given a fixed budget. One tries to maximize the cumulative rewards over all allocations. This perspective is also taken in response-adaptive trials \citep{rosenberger1993use, hu2006theory}, which allocate more patients to the better treatment based on the available patient responses. Thus, such trials can be viewed as exploitative, since most of them adopt myopic strategies by exploiting the current best arm. Our method is more exploratory, exploring patients close to the decision boundary to learn better individualized treatment rules for {\em future} practice. Some adaptive enrichment designs \citep{wang2009adaptive,simon2013adaptive} are similar in that they allow the eligibility criteria to be updated during the trial. However, our primary goal is to construct informative and favorable ITRs rather than to confirm the efficacy of one treatment over another. Our method is also different from adaptive randomization procedures, which primarily aim to place more subjects onto the treatment arm that is more promising \citep{rosenberger2012adaptive}.  \citet{deng2011active} proposed a minimax bandit model for clinical trials that carefully distributes the trial resources, with the same objective as our method. However, they specify subpopulations in advance, and the optimal treatments are learned for each subpopulation. The resulting ITRs may not be ideal if the subpopulations are not formed correctly.

In Section \ref{scn_method}, we introduce a general methodology for conducting active clinical trials, along with methods for constructing optimal ITRs using the accumulated data. In Section \ref{scn_theory}, we discuss the theoretical properties of our approach by providing a finite sample upper bound on the difference in the expected outcome under the estimated ITR and the optimal ITR. In
Section \ref{scn_simulation}, we conduct extensive simulation studies to examine the empirical performance and also compare the results with those of \citet{deng2011active}. Two real data examples are presented in Section \ref{scn_data}, and Section \ref{scn_discussion} provides a discussion. The technical details are presented in the Appendix and the Supplementary Materials.

\section{Methodology} \label{scn_method}

\subsection{Optimal Individual Treatment Rule}\label{sec:review}

In this section, we discuss a probabilistic framework for studying the optimal ITRs that is similar to that of \cite{qian:itr11} and \cite{Zhao:OWL12}.
Let $(\BX,A,R)$ be a random triple with a joint distribution $P$.
Here, $\BX \in \mb R^p$ denotes the patient's baseline covariates with marginal distribution $\Pi$, $A$ is a binary treatment assignment taking values in $\{-1, 1\}$, and $R$ stands for the treatment outcome (a larger value of $R$ corresponds to a better outcome). An ITR $D(\cdot)$ is defined as a function from the covariate space $\mb R^p$ into the treatment space $\{-1,1\}$.

We use the value function, denoted as $V(D)$, to measure the quality of $D$, which is a marginal mean outcome representing the overall population mean were all patients to receive treatment according to $D$. Our goal is to identify the optimal ITR that yields the maximum $V(D)$.
For any ITR $D$, let $P^{D}$ be the distribution of $(\BX,A,R)$ when $A=D(\BX)$, and let $\mb E^{D}$ be the corresponding expectation.
The value function $V(D)=\mb E^{D}(R)$.
\cite{qian:itr11} show that $V(D)$ can be expressed as
\begin{equation}
V(D)=\mb E\left[\frac{RI(A=D(\BX))}{\pi(A;\BX)}\right], \label{valuefcn}
\end{equation}
where $\mb E$ denotes the expectation w.r.t. the joint distribution $P$, $I(\cdot)$ is an indicator function, and $\pi(a;\BX)$ is the
conditional probability $P(A=a|\BX)$ for $a\in \{-1,1\}$.
For simplicity, we assume a pure randomization scheme with equal probability for different assignments, i.e., $\pi(a;\BX)=1/2$, throughout the paper.

Let $D^*$ denote the optimal treatment rule that maximizes $V(\D)$.
By rewriting $V(D)$ as $V(D)=\mb E\Big(\mb E[RI(D(X)=1)|A=1,X]+\mb E[RI(D(X)=-1)|A=-1,X]\Big)$,
we obtain
\begin{equation}
D^*(\Bx)=\sign\{f^*(x)\},
\label{optdecision}
\end{equation}
where $f^*(x):=\mb E[R|A=1,\BX=\Bx] -\mb E[R|A=-1,\BX=\Bx]$ is called the \textit{contrast function}.
The \textit{optimal decision boundary} is just the level set
$\{\Bx\in \mb R^p: f^*(\Bx)=0\}$. As can be seen, the overall benefit will be maximized if the patients satisfying $f^*(x) \geq 0$ receive the alternative treatment (treatment 1), and the others receive standard care (treatment -1). Hence, the decision is tailored according to the patient's characteristics represented by $X$.

To estimate the optimal ITR $D^*$ from the data $\{X^{(i)},A^{(i)},R^{(i)}\}_{i=1}^{n}$, usually collected from a clinical trial, one can fit a parametric or nonparametric regression model for $\mb E[R|A=a,\BX=x]$ (equivalently, $f^\ast(x)$), and then estimate the optimal ITR by plugging the fitted model into (\ref{optdecision}). Alternatively, we can replace the problem of maximizing (\ref{valuefcn}) by minimizing a weighted classification error $\mb E\left[{RI(A\neq{D}(\BX))}/{\pi(A;\BX)}\right]$.
In this case, existing classification techniques, e.g., support vector machines, can be adapted to estimate $D^*(\Bx)$; see \citet{Zhao:OWL12} for more details. All of the above methods take the whole (randomized) clinical trial data as an input, and thus they have no influence on the data collection process. They belong to the class of \textit{passive learning} methods, also called \textit{batch learning}.
In clinical trials, the patient recruiting process is usually long, and the treatment and monitoring process can be extremely expensive. With a limited budget and a fixed sample size, we should wisely allocate the resources, i.e., decide who to recruit, in order to learn the optimal ITRs at a lower cost. This motivates us to propose active clinical trials that can identify the optimal ITRs with a significantly reduced cost.

\subsection{Active Clinical Trials}

The active learning (AL) approach in the classification literature is shown to produce accurate classifiers with a significantly reduced number of label requests; see \citet{Balcan2008The-True-Sample00}, \citet{dasgupta2007general}, \citet{castro2008minimax}, \citet{koltchinskii2010rademacher}, \citet{hanneke2011rates}, and \citet{minsker2012plugin}. Recall that the estimation of ITRs can be thought of as a classification problem. For example, a patient with a small outcome given the assigned treatment is potentially misclassified. We use AL techniques to select the ``most informative'' patients based on a given pool of prognostic variables, and then we randomize only the patients with these ``informative'' characteristics. The intuition behind this patient-selection process is simple: if we have confidence that certain patients will benefit from a particular treatment, we should not recruit them into the study since  they are not likely to contribute to the estimation of ITRs. However, it is hard to determine the optimal treatment, i.e., the sign of $f^*(x)$, for the patients whose baseline variables are ``close'' to the optimal decision boundary.
As will be seen from the empirical and theoretical analysis, patients with such features are more ``informative'' in the sense that their outcomes after randomization provide more insight into the optimal ITRs (to be applied to {\em future} patients).
Moreover, since patients will not be randomized if we have confidence in the treatment they should receive, they are less likely to be exposed to ineffective medications.

We assume that there is no delay in observing the outcome. For $S\subset \supp(\Pi)$ and a function $f:\mb R^p\mapsto \mb R$, let $f|_S:S\mapsto \mb R$ be the restriction of $f$ onto $S$,
and define $\|f\|_{\infty,S}=\left\|f|_S\right\|_\infty:=\sup\limits_{x\in S}|f(x)|$.
Given $\delta>0$, set
\[
\m F_{\infty,S}(f,\delta):=\{g:S\mapsto \mb R: \ \|g-f\|_{\infty,S}\leq \delta\}
\]
as a $\delta$-band around $f$ on $S$. A precise description is given in Algorithm~\ref{alg1}. Here, we introduce the important notion of an \textit{active set} \citep{minsker2012plugin}, which underlies the majority of active learning algorithms. At step $k$ of our Algorithm \ref{alg1}, the active set $AS_k$ is defined to be the set of baseline variables for which the best treatment is not yet known. The active set is characterized by the confidence interval, i.e., $x$ belongs to the active set if and only if the confidence interval for $f^\ast(x)$ contains both positive and negative elements; Figures  \ref{fig:01} and \ref{fig:02} below illustrate this. The approximation of $AS_k$ using a regular set $\act_k$ is discussed in Appendix  A.1.

\begin{algorithm}[h]
{\scriptsize
\SetKwInOut{Input}{input}\SetKwInOut{Output}{output}
\Input{Sample size limit $N$; confidence $\alpha$}
\Output{$\widehat D:=\sign(\widehat f)$}
\nl $k=0$, $\act_0:=\supp(\Pi)$\;
Set the initial $N_0:=2\lfloor\sqrt{N}\rfloor$, \label{b1} and $LB:=N-N_0$\;
\For{$i=1$ \KwTo $N_0$}
{
 Recruit $X^{(i,0)}$ from $\Pi$\;
 Randomize $X^{(i,0)}$ to treatment $A^{(i,0)}=1$ or $-1$ with equal probability\;
 Observe $R^{(i,0)}$\;
}
\nl  Construct the estimator $\widehat f_0(x)$ of $f^*$ from $S_0=\l\{\left(X^{(i,0)},A^{(i,0)},R^{(i,0)}\right)\r\}_{i=1}^{N_0}$\;
\While{$LB > 0$}
{
\nl $\widehat{\m F}_k:=\left\{f\ : \ f|_{{\act}_k}\in\mathcal{F}_{\infty,\act_k}(\widehat f_k;\frac{3}{2}\delta_{k}), \
 \ f|_{\supp(\Pi)\setminus \act_k}\equiv\widehat f_{k-1}|_{\supp(\Pi)\setminus \act_k}\right\}$  \tcc*[f]{$\delta_k$: confidence band size}\;
\nl  $k:=k+1$\;
$AS_k:=\left\{x\in \supp(\Pi): \ \exists f_1,f_2\in \widehat{\mathcal{F}}_{k-1}, \sign(f_1(x))\ne\sign(f_2(x))\right\}$\tcc*[f]{active set}\;
Approximate $AS_k$ with a regular set $\act_k$\;
\nl  \If{$\act_k\cap \supp(\Pi)=\emptyset$}{{\bf break}}
 \Else{
$N_k=2N_{k-1}$\;
\For{$i=1$ \KwTo $\lfloor N_{k}\cdot \Pi(\act_k)\rfloor$}
 {
Recruit $X^{(i,k)}$ from the active set $\widehat\Pi_k:=\Pi_{\act_k}(dx)$\;
Randomize $X^{(i,k)}$ to treatment $A^{(i,k)}=1$ or $-1$ with equal probability\;
Observe $R^{(i,k)}$, $S_k:=\l\{\left(X^{(i,k)},A^{(i,k)},R^{(i,k)}\right), \ i\leq \lfloor N_{k}\cdot \Pi(\act_k)\rfloor\r\}$\;
}
Construct the estimator $\widehat f_k(\cdot)$ of $f^*$ based on $S_k$\;
$LB:=LB- \lfloor N_{k}\cdot \Pi(\act_k)\rfloor$\;
$\widehat f:=\widehat f_k$ \tcc*[f]{keeping track of the most recent estimator}\;
 }   }
 }
\caption{Active Clinical Trials for Personalized Medicine}
\label{alg1}
\end{algorithm}

\begin{figure}[ht]
\begin{minipage}[b]{0.45\linewidth}
\centering
\includegraphics[width=\textwidth,height=4cm]{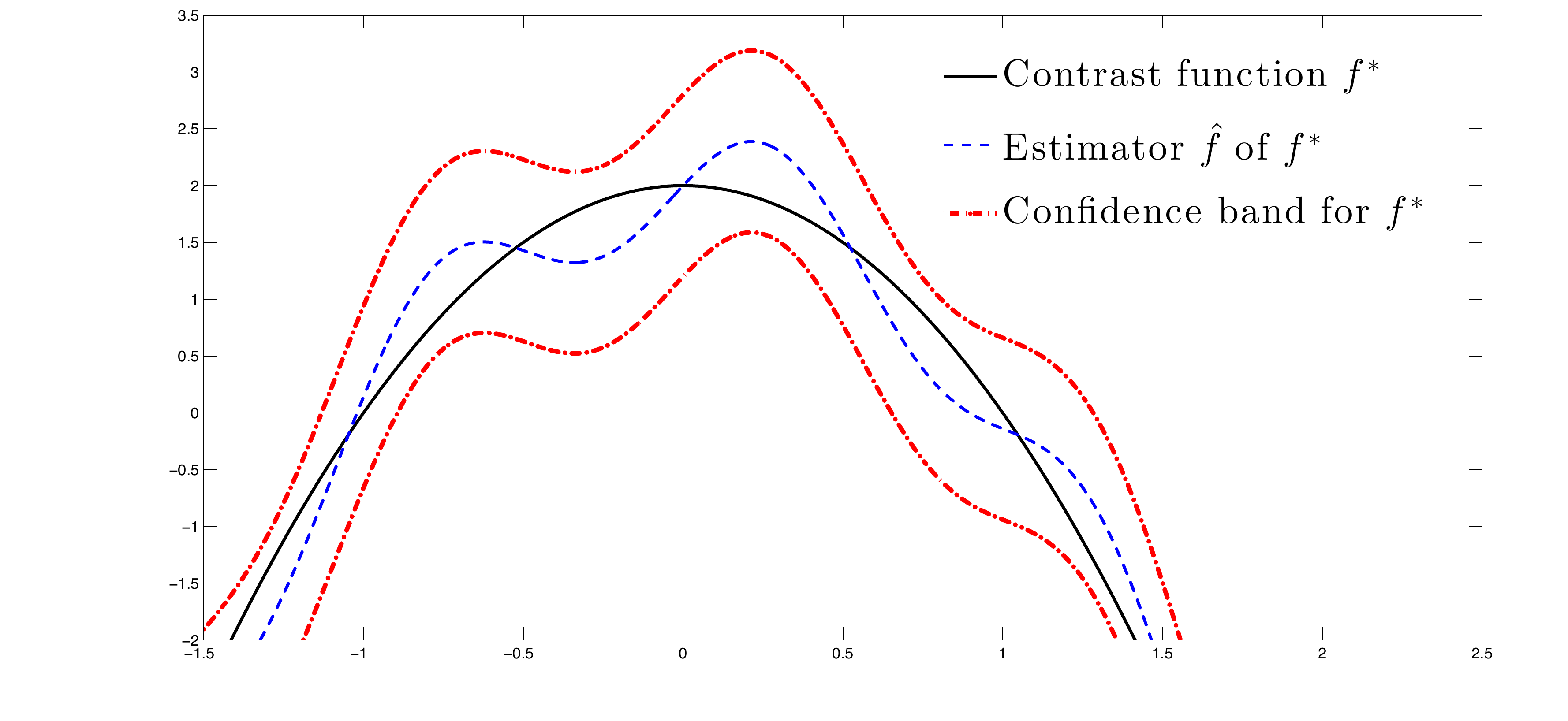}
\caption{Confidence band for the contrast function.}
\label{fig:01}
\end{minipage}
\begin{minipage}{0.08\linewidth}
\text{}
\end{minipage}
\begin{minipage}[b]{0.45\linewidth}
\centering
\includegraphics[width=\textwidth,height=4cm]{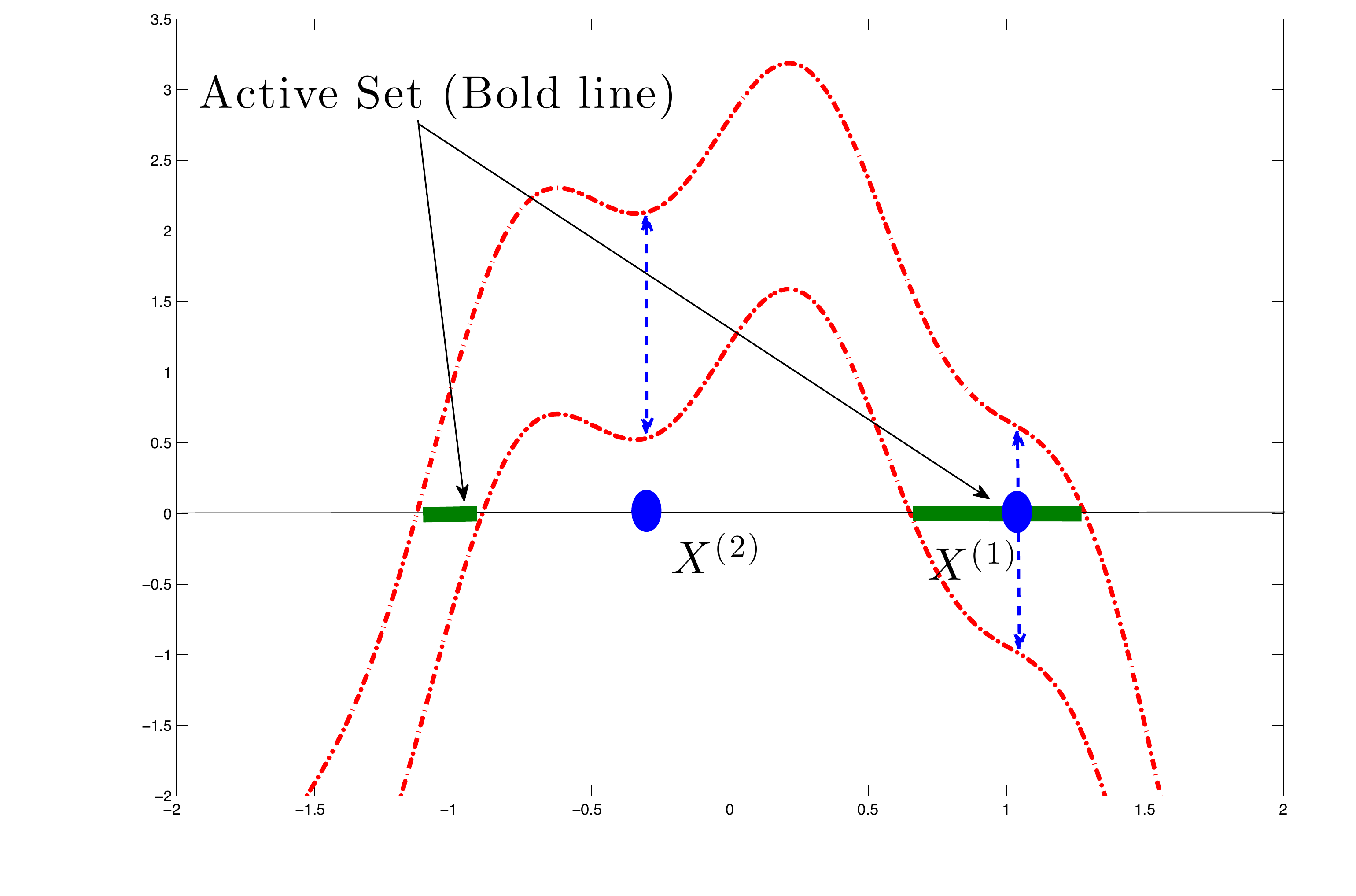}
\caption{$X^{(1)}$ belongs to the active set; $X^{(2)}$ does not.}
\label{fig:02}
\end{minipage}
\end{figure}

In practice, it is not necessary to evaluate the active sets explicitly at each step. Instead, at step $k$, we repeat the following two steps until we reach the number of patients that can be randomized:
\begin{enumerate}
\item Recruit a group of new patients with baseline variables $\BX^{(i,k)}, i=1,\ldots,N_k$.
\item Let $\widehat f_{i,k}(X^{(i,k)})$ be the estimator of $f^\ast(X^{(i,k)})$ based on $S_k$ and $I_{i,k}=[\widehat f_k(X^{(i,k)})-\delta(X^{(i,k)}),\widehat f_k(X^{(i,k)})+\delta(X^{(i,k)})]$ be the confidence interval, where $S_k$ is defined in Algorithm 1 and represents the available sample in the $k^{th}$ step.
\begin{enumerate}
\item If $0 \notin I_{i,k}$: drop the patient from the study.
\item If $0 \in I_{i,k}$: randomize the patient, record the outcome $R$, and add $(X^{(i,k)},A^{(i,k)},R^{(i,k)})$ to the current sample.
\end{enumerate}
\end{enumerate}
\noindent The above procedure gives a data set containing ``the most informative'' observations for predicting the optimal treatment. For every new patient, this data set will be used to predict the treatment rule based on that individual's baseline variables. At each iteration, we must specify an estimator of the contrast function and the corresponding confidence interval. In the next section, we will propose two construction methods based on kernel estimation and Gaussian process regression, respectively. We remark that the empirical performance of our method is essentially controlled by the confidence level that we set. This level also determines the rejection ratio, e.g., $N=100$ out of 140 total iterations in simulation Scenario 1 of Section \ref{scn_simulation}.

\subsection{Confidence Interval Construction}
\label{scn_act}

\subsubsection{Kernel Smoothing Approach}
\label{scn_kernel}

Let $K$ be a smooth kernel function (with bandwidth $h_n$) satisfying Assumption (A2) in Section \ref{scn_theory}.
Define $K_{h_n}(\bx_0-\bX):=K((\bx_0-\bX)/h_n)/h_n^p$.
Let $\eta_j(x):=\mb E[R|A=j, X=x], j=\pm 1,$ be the conditional expectation of $R$ given $A$ and $X$.
Given $n$ observations from $P$, we propose the estimators $\widehat\eta_j(\cdot;h_n)$ for $j=\pm 1$, and the plug-in estimator $\widehat f(\cdot;h_n)$ of the contrast function $f^\ast(\cdot)$.
At any fixed point $\bx_0$,
\begin{align}
\label{fhat}
&
\widehat \eta_j(x_0;h_n)=\frac{\sum\limits_{i=1}^{n} K_{h_{n}}(\bx_0-\bX^{(i)})I(A^{(i)}=j) R^{(i)}}{\sum\limits_{i=1}^{n} K_{h_{n}}(\bx_0-\bX^{(i)})I(A^{(i)}=j)}, \ j=\pm 1, \\
& \nonumber
\widehat f(\bx_0;h_n)=\widehat \eta_1(x_0;h_n)-\widehat \eta_{-1}(x_0;h_n).
\end{align}
Here, $h_n = h_n(\bx_0)$ is an adaptive bandwidth parameter varying with $\bx_0$. We need to choose a proper $h_n(\bx_0)$, which controls the local amount of data near $\bx_0$, to optimally balance the estimation bias and variance; see Appendix~\ref{app:ker} for the technical details.

Assuming that both $\eta_1$ and $\eta_{-1}$ are Lipschitz continuous with Lipschitz constants bounded by $L$,
we define
\begin{align}
\nonumber
h_{n,j}(x_0)=\inf\l\{h>0: \ L^2 h^2\geq \frac{C_1(K,P)}{\sum\limits_{i=1}^n I\l\{\|x_0-X^{(i)}\|_2\leq h\r\}I\{A^{(i)}=j\}} \r\}, \ j=\pm1,
\end{align}
where $C_1(K,P)$ is a constant depending on the kernel $K$ and distribution $P$, and $\|\cdot\|_2$ is the usual Euclidean norm. Set
\begin{align}
\label{eq:b-v}
h_n(x_0):=\max\l(h_{n,1}(x_0),h_{n,-1}(x_0)\r).
\end{align}
This choice mimics the usual ``bias-variance tradeoff'': indeed,
the bias of $\widehat \eta_j(x_0;h)$ is bounded by the order of $L h$, while
$\Big(\sum\limits_{i=1}^n I\l\{\|x_0-X^{(i)}\|_2\leq h\r\} I\{A^{(i)}=j\}\Big)^{-1}$
plays the role of the variance parameter. Based on the above choice of $h_n(\bx_0)$, i.e., (\ref{eq:b-v}), we define the radius of the confidence interval to be
\begin{equation}
\delta(\bx_0):=t \cdot L h_n(\bx_0),
\nonumber
\end{equation}
where $t$ controls the coverage probability. The display above depends on the unknown constants $C_1(K,P)$ that must be chosen before running the algorithm. Given a certain confidence level, we recommend selecting the ``confidence parameter'' $t$ by reverting the coverage probability error that decays exponentially fast with $t$; see Corollary~\ref{cor:1}.

\noindent{\bf Remark 1.} {\it We are aware that the proposed kernel methods are unfortunately affected by the ``curse of dimensionality.'' However, our theoretical analysis reveals that the empirical performance of our kernel estimate essentially depends on the ``intrinsic dimension'' $d$ of the support of the marginal distribution $\Pi$ (see Assumption (A3) in Section~\ref{scn_theory}), which might be much lower than the ambient dimension $p$. The concept of the intrinsic dimension originated in geometry and was later introduced to statistics, e.g., \cite{allard2012multi}. It characterizes a low-dimensional representation embedded in a high-dimensional space. For example, when the data are linear, we may use principal component analysis to identify the subspace that contains the data. The number of important components is the intrinsic dimension $d$, which can be significantly smaller than the total number of covariates $p$. In practice, we follow the local singular value decomposition method, suggested in \citet{little2009multiscale}, to determine the intrinsic dimension. }

\subsubsection{Gaussian Process Regression Approach}
\label{scn_bayes}

The new method presented in this section is particularly well suited for applications since it is completely data-driven and does not require us to specify any unknown parameters in advance; its performance is demonstrated in Section \ref{scn_simulation}. However, the price is that we must assume that the conditional distribution of the outcome $R$ given $(A,X)$ is Gaussian. In other words, if an individual with baseline variables $X$ receives treatment $A=j$, we assume that the outcome satisfies $R=\eta_{j}(\bX)+\eps$ for $j=\pm 1$, where $\eps\sim N(0,\sigma^2)$. In this situation, we take a Bayesian approach by imposing a Gaussian process prior (defined by the mean $m(\cdot)$ and covariance $k(\cdot,\cdot)$) on functions $\eta_1(\cdot)$ and $\eta_{-1}(\cdot)$; see Chapter 2 of \citet{Rasmussen2006Gaussian-proces00}. For simplicity, the mean function $m(\cdot)$ is set to be identically zero throughout this paper.

As before, the confidence interval of the contrast function builds upon those of $\eta_{1}(\cdot)$ and $\eta_{-1}(\cdot)$. Hence, we start from the inference procedure for $\eta_1(\cdot)$. Let $\{(X^{(i)},R^{(i)})\}_{i=1}^{n_1}$ be the observations corresponding to the individuals who received treatment $1$. The covariance function of the Gaussian process prior is set as a slight variant of the squared exponential kernel: given $x=(x_1,\ldots,x_p)$ and $x'=(x'_1,\ldots,x'_p)\in \mb R^p$,
\[
k_\gamma(x,x')=\gamma_0\exp\left(-\sum_{l=1}^p \frac{(x_l-x'_l)^2}{2\gamma_l^2}\right)
\]
for some positive $\gamma_0,\ldots,\gamma_p$.
Let $K_\gamma$ be a $p\times p$ matrix with entries $(K_\gamma)_{l_1,l_2}=k_\gamma(X_{l_1},X_{l_2})$, $l_1, l_2 = 1,\ldots,p$.
Under the Gaussian error assumption, the marginal distribution of the vector $\textbf{R}=(R^{(1)},\ldots,R^{(n_1)})^T$ given $\textbf{X}=(X^{(1)},\ldots,X^{(n_1)})$ is a multivariate Gaussian with mean $0$ and covariance matrix
$
K_{\textbf{R}}:=K_\gamma+\sigma^2 I_p,
$
where $I_p$ is a $p\times p$ identity matrix. The ``optimal'' value $\Gamma_\ast:=(\gamma_0^\ast,\ldots,\gamma_p^\ast,\sigma_\ast^2)$ of the parameter $\Gamma=(\gamma_0,\ldots,\gamma_p,\sigma^2)$ is then ``learned'' from the data by finding a local maximum of the marginal log-likelihood
\[
\log p(\textbf{R}|\textbf{X},\Gamma):=-\frac{1}{2} \textbf{R}^T K_{\textbf{R}}^{-1}\textbf{R} -\frac{1}{2}\log\det K_{\textbf{R}}-\frac{n_1}{2}\log 2\pi
\]
with respect to $(\gamma_0,\ldots,\gamma_p,\sigma^2)$; see Section 5.4.1 in \citet{Rasmussen2006Gaussian-proces00} for more details. Therefore, the Gaussian process regression is a {\em global} method that can automatically select the bandwidth by maximizing the data likelihood. This is in contrast with the previous kernel method where the bandwidth is selected {\em locally}; see Equation (\ref{eq:b-v}) above.

We next construct the confidence interval based on the posterior distribution of $\eta_1(\cdot)$ with the optimal $\Gamma_\ast$. Given a new observation with the baseline variable $x_0$, the value of $\eta_1(x_0)$ is estimated by the posterior mean, denoted $\widehat \eta_1(x_0)$, i.e.,
\begin{align}
\nonumber
\widehat \eta_1(x_0):=k_{\Gamma_\ast}(x_0,\textbf{X})\,K_{\textbf{R}}^{-1}\textbf{R},
\end{align}
where $k_{\Gamma_\ast}(x,x'):=\gamma_0^\ast \exp (-\sum\limits_{l=1}^p  {(x_l-x'_l)^2}/{2(\gamma^\ast_l)^2})$ for $x,x'\in \mb R^p$, and
\[
k_{\Gamma_\ast}(x_0,\textbf{X})=(k_{\Gamma_\ast}(x_0,X^{(1)}),\ldots,k_{\Gamma_\ast}(x_0,X^{(n_1)})).
\]
The variance of the posterior distribution is given by
\begin{align}
\nonumber
\widehat{\sigma}_1^2(x_0):=k_{\Gamma_\ast}(x_0,x_0)-k_{\Gamma_\ast}(x_0,\textbf{X})\,K_{\textbf{R}}^{-1}\,k_{\Gamma_\ast}(x_0,\textbf{X})^T.
\end{align}
The square root of the posterior variance naturally controls the length of the confidence interval for $\eta_1(x_0)$.
For example, setting $\delta_1(x_0):=3\widehat{\sigma_1}(x_0)$ gives the confidence interval for $\eta_1(x_0)$ of  posterior probability $>99\%$ (note that this may not correspond to the frequentist coverage).

Define $\widehat \eta_{-1}(x_0)$ and the associated $\delta_{-1}(x_0)$ analogously. We thus obtain
a confidence interval with the center
$\widehat f(x_0):=\widehat \eta_{1}(x_0)-\widehat \eta_{-1}(x_0)$ and the radius
$\delta(x_0):=\delta_1(x_0)+\delta_{-1}(x_0)$.
The numerical implementation of this Bayesian inference procedure can easily be performed with the \textbf{gpml} Matlab toolbox \citep{Rasmussen2010Gaussian-proces00}.

\section{Theoretical Analysis}\label{scn_theory}

In this section, we focus our theoretical analysis on the kernel smoothing approach. The frequentist property of the credible set for the nonparametric Bayesian method has not been well developed. Hence, a theoretical analysis of the Gaussian process regression method is beyond the scope of this paper.

For simplicity, we suppose that the marginal distribution $\Pi$ of the baseline variable vector $X$ is known. Note that Algorithm \ref{alg1} does not need to know or estimate $\Pi$ explicitly, and this assumption is only for the theoretical analysis. In addition, we assume the following conditions on the kernel $K:\mb R^p\mapsto \mb R$ and the distribution $\Pi$:

\begin{description}
\item[(A0)]
Both $\eta_1(x):=\mb E[R|A=1,\BX=x]$ and $\eta_{-1}(x):=\mb E[R|A=-1,\BX=x]$ are Lipschitz continuous on $\mb R^p$ with Lipschitz constants bounded by $L$.
\item[(A1)] The random variable $|R|$ is bounded by $0<M<\infty$ a.s.
\item[(A2)] $K(x)$ is a nonnegative, compactly supported, Lipschitz-continuous function with a Lipschitz constant $L_K$. Moreover,
$K(x)\geq \ell_K I\{\|x\|_2\leq 1\}$ for some $\ell_K>0$.
\item[(A3)] The ``intrinsic dimension'' of $\supp(\Pi)$ is equal to $d$ for some integer $d\leq p$, and $\Pi$ is equivalent to the uniform distribution over its support; see Appendix \ref{asm:a3} for a more precise statement.
\item[(A4)] Margin condition: there exist $K_2=K_2(\Pi),\gamma=\gamma(\Pi)>0$ such that for all $t>0$
\[
\Pi\l(x: \ |f^*(x)|\leq t\r)\leq K_2 t^\gamma.
\]
\end{description}

Assumption (A3) says that over a ``nice'' set, $\Pi$ is close to the uniform distribution; see condition (\ref{ass:int}) in the Appendix.
The intrinsic dimension $d$ is crucial here: in many applications, $p$ is large but $d$ is small (see Remark 1).
Note that the rate in our main result, Theorem \ref{thm:conv}, depends on $d$ but not on $p$.
Assumption (A4) is an analogue of the well-known \textit{margin condition} \citep{tsybakov2004optimal}, which is commonly used to characterize the complexity of a binary classification problem. Larger values of $\gamma$ mean that the two treatment effects are less likely to be similar, yet in nontrivial examples, $\gamma\in [0,d]$.
In particular, as indicated in Proposition 3.4 in \cite{Audibert2007Fast-learning-r00}, for a smooth contrast function $f^*(x)$, $\gamma \in [0,d]$ unless $f^*(x)$ does not cross $0$ at any point in the interior of \supp($\Pi$), i.e., all the patients benefit from one treatment. Note that our analysis does not require $\gamma$ to be known in advance.

We are now ready to present the (finite-sample) performance guarantee for our method.
\begin{thm}
\label{thm:conv}
Let $h_k =[\{\log(N/\alpha)+d\log(N_k)\}/{N_k}]^{1/(d+2)}, k=0,\ldots, L$, where $L$ is the total number of  iterations in Algorithm~\ref{alg1}. Set the associated $\delta_k=4Ch_k$, where $C$ is a constant specified in Lemma \ref{lem:sup-norm}.
With probability greater than $1-\alpha$, the estimator $\widehat D$ of $D^\ast$ returned by Algorithm~\ref{alg1} satisfies
\[
\left|V(\widehat D) - V(D^\ast)\right| \leq \widetilde C N^{-\frac{1+\gamma}{2+d-\gamma}}\l(\log(N/\alpha)\r)^{\theta},
\]
where $N$ is the number of randomized subjects, $\theta=\frac{(4+2d-\gamma)(1+\gamma)}{(2+d)(2+d-\gamma)}$, and $\widetilde C$ is a constant that depends on the kernel $K$ and distribution $\Pi$.
\end{thm}

It is worth noting that when $\gamma$ is large (say, close to $d$), the rate of Theorem \ref{thm:conv} is ``almost'' dimension-free. We also remark that  the rate of Theorem \ref{thm:conv} can not be uniformly improved by any active learning technique, as shown in \cite{minsker2012plugin}.
\citet{qian:itr11} used a {\em parametric} modeling approach, i.e., they fitted an $L_1$-penalized regression model to estimate the optimal ITRs, and they obtained a rate of $(\log N/N)^{(1+\gamma)/(2+\gamma)}$ with an appropriate choice of tuning parameter.
However, their model could be misspecified in practice. In contrast, our method is {\em nonparametric} with possibly minimal model assumptions.

\section{Simulation Studies}
 \label{scn_simulation}

In this section, we assess the empirical performance of the active clinical trial method.
Let $X = (X_1,X_2, \ldots, X_p)$, where $X_1,\ldots, X_p$ are independent of each other.
The distribution of $X$ varies according to different scenarios detailed below.
The treatment $A$ is generated from $\{-1,1\}$ with equal probability.
The response $R$ is generated from $N(Q_0(X,A), 1)$, where
$$Q_0(X,A) = m_0(X) + T_0(X,A).$$
Here, $T_0(X,A)$ is the interaction between the treatment and the baseline variables.
In what follows, $U[a,b]$ stands for the uniform distribution on the interval $[a,b]\subset \mb R$.
Consider four scenarios for $T_0(X,A)$:

\begin{enumerate}
 \item $X_l\sim U[-1,1], l=1,2, m_0(X)=1+2X_1+X_2, T_0(X,A) = 0.5(1-X_1-X_2)A.$
  \item $X_l\sim U[-1,1], l=1,2, m_0(X)=1+2X_1+X_2, T_0(X,A) = 1/2+(1-X_1^2-X_2^2)(X_1^2+X_2^2-1)A.$
  \item $p=3, m_0(X) = 1+2X_1+X_2-X_3, T_0(X,A)= 1.5(X_1X_2(1+X_3))A$, where $X_1,X_2,X_3$ are on the sphere generated as follows. Let $\widetilde{X}_1,\ldots,\widetilde{X}_3 \sim U[-1,1]$, and
  $$X_l=\frac{\widetilde{X}_l}{\sqrt{\sum_{l=1}^3 \widetilde X_l^2}}I\Big(\sum_{l=1}^3 \widetilde X_l^2\leq 1\Big), l=1,2,3.$$
  \item $p=8, m_0(X) = 1+2X_1+X_2-X_3, T_0(X)=0.2(\sum_{l \text{ is even}}X_l-\sum_{l \text{ is odd}}X_l)A$, where $X_1,\ldots,X_8$ have uniform distribution $U[-1,1]$.
    \end{enumerate}

It can be seen that the optimal ITR for Scenario 1 is linear, i.e., $\D^*(X)=\sign(1-X_1-X_2)$.  The optimal ITR for Scenario 2 is nonlinear,  with $D^*(X) = I (0.3 \leq X_1^2+X_2^2 \leq 1.7)$. Scenario 3 represents the case where the data are supported on the manifold, i.e., a two-dimensional sphere in $\mb R^3$. The treatment effect in Scenario 3 is highly nonlinear with $D^*(X)=\sign(X_1X_2(1+X_3))$. Scenario 4 has a relatively high dimensional covariate, i.e., 8, with a linear treatment effect.

We apply the active clinical trial outlined in Algorithm \ref{alg1}.
We implemented both the kernel method in Section \ref{scn_kernel}, denoted AL-BV, and the Gaussian process regression method in Section \ref{scn_bayes}, denoted AL-GP. We used the \textbf{gpml} Matlab toolbox \citep{Rasmussen2010Gaussian-proces00} in the latter method. The active clinical trial proceeds by selectively recruiting subjects whose differential treatment effects are smaller than a threshold, i.e., Steps 3--4 of Algorithm \ref{alg1}. This iterative procedure
screens out a certain number of subjects, whose optimal treatments can be determined with high confidence, and retains the remaining $N$ subjects for estimating the optimal ITR.
The active learning algorithm proposed in \citet{deng2011activedeveloping} serves as a comparison. The designed trial therein  focuses on simple ITRs that utilize a small number of subpopulation categories to personalize the treatment. In particular, they assume the subpopulations are known a priori. Throughout the clinical trial, they sequentially select the (subpopulation, treatment) pair so as to minimize the maximal error of selecting a suboptimal treatment. For all the scenarios, we form four subgroups by dichotomizing $X_1$ and $X_2$ at 0.  We use MINMAXPICS, the algorithm name in their paper, to denote this method.

In addition, we compare with two passive learning approaches: the outcome weighted learning (OWL) method \citep{Zhao:OWL12} and the ordinary least square (OLS) method. Both methods recruit subjects upon arrival in the clinical trial and estimate the optimal ITR using the collected data after the trial ends.
$N$ subjects are randomly selected for both methods. In the OWL method, using the available data, we minimize the target function
$
 \pn\left[{R\phi(Af(\bX))}/{\pi(A;\BX)}\right] + \lambda_n\|f\|_2^2,
$
where $\phi(t)=\max(1-t,0)$ is the hinge loss, $\pn$ is the empirical measure, and $\lambda_n$ is a tuning parameter controlling the amount of penalization. We consider a nonlinear functional space for the ITRs, and a Gaussian kernel is used in the implementation. The optimal ITR is estimated via $\widehat\D(\Bx) = \sign(\widehat f(X))$, where $\widehat f(x)$ are the minimizers of the above objective function. In the OLS method, we first regress $R$ on $(\BX, A, \BX A)$, and then estimate the optimal ITR by finding the treatment that yields a larger predicted outcome for each individual.

The initial sample size $N_0$ is fixed at $50$, while the additional sample size $N-N_0$ is $50, 100, 200, 300, 400, 500$, or $800$.
To evaluate the empirical performance, we generated a testing dataset of sample size $10000$, mimicking a large pool of future subjects.
The estimated ITRs  $\widehat{D}(X)$ using the different methods are validated on this large testing set.
Since the main effect is invariant across different ITRs, we can calculate the average excess value $AEV(D^*,\widehat{D})$ as
$$AEV(D^*,\widehat{D})=\frac{1}{n}\sum_{i=1}^{n}[T_0(X^{(i)},D^*)-T_0(X^{(i)},\widehat{D})],\,\, n=10000$$
where the empirical average is taken over the validation set.
This quantity directly reflects the expected clinical benefits for future subjects treated according to $\widehat{D}(X)$, with a smaller value indicating a better treatment decision.
We repeat the process $1000$ times and average the resulting values over all the runs. In Figure~\ref{sim_fig}, we plot  $\log\{AEV(D^*,\widehat D)\}$ against $\log(N-N_0)$, where $\widehat D$ was obtained using each method. The log-scale is used to give a better display of the polynomial convergence rates in the different methods.

In all the scenarios, our active clinical trials perform uniformly better than OWL. In Scenario 1, the treatment effect is linear, which indicates that OLS is the best possible method. However, AL-GP has comparable performance, especially when the sample size is large.
The performance of AL-BV also improves with $N$. When the treatment effect is nonlinear, as in Scenarios 2 and 3, the strength of active learning is clearly demonstrated.
Both methods initially perform better for small sample sizes and then converge much faster as the sample size grows.
In contrast, the values of the estimated ITRs from the other two methods do not converge (Scenario 3) or do not converge to the optimal value (Scenario 2). In Scenario 4, the number of covariates is increased to eight. This places severe difficulties on the kernel estimation because of the curse of dimensionality. However, with the linear treatment effect, the AL-GP results are satisfactory compared with those of OLS. The MINIMAXPICS algorithm performs the best in Scenario 3, where the subpopulations are correctly defined in advance. When the prespecification of the subpopulations is incorrect, each subpopulation contains both patients who benefit from treatment 1 and patients who benefit from treatment -1. In this case, the constructed ITRs are not ideal, even with large sample sizes. In general, we believe that active learning methods provide robust results for various treatment mechanisms, which are usually unknown in practice. We presented some additional simulation results in the supplementary materials, where we investigated scenarios with biomarkers of normal distribution, and conducted several sensitivity analyses.

\begin{figure}[h!]
\caption{Excess values (log scale). Initial size was set to 50.}
\label{sim_fig}
\begin{center}
\includegraphics[width=3.2in,height=3.2in]{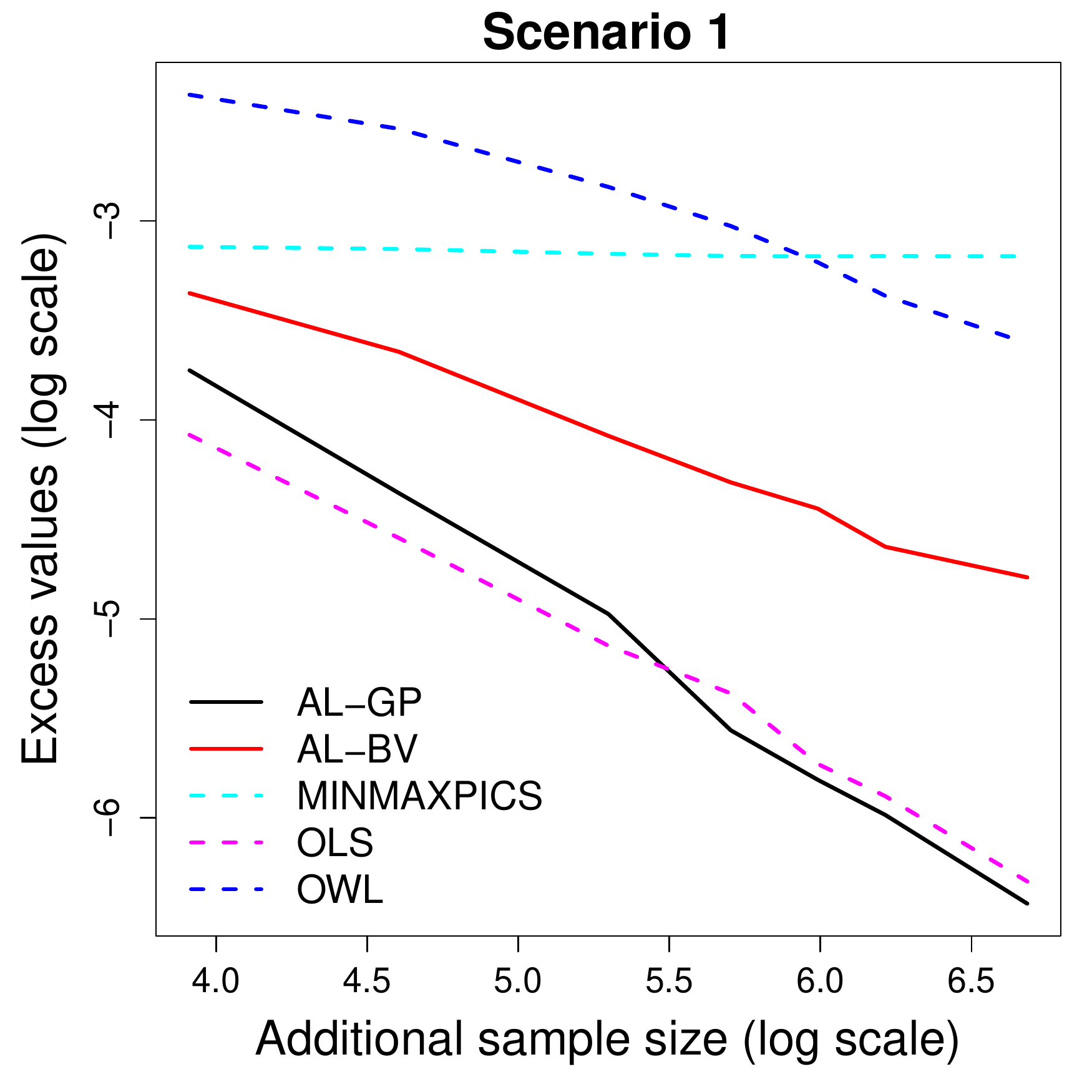}
\includegraphics[width=3.2in,height=3.2in]{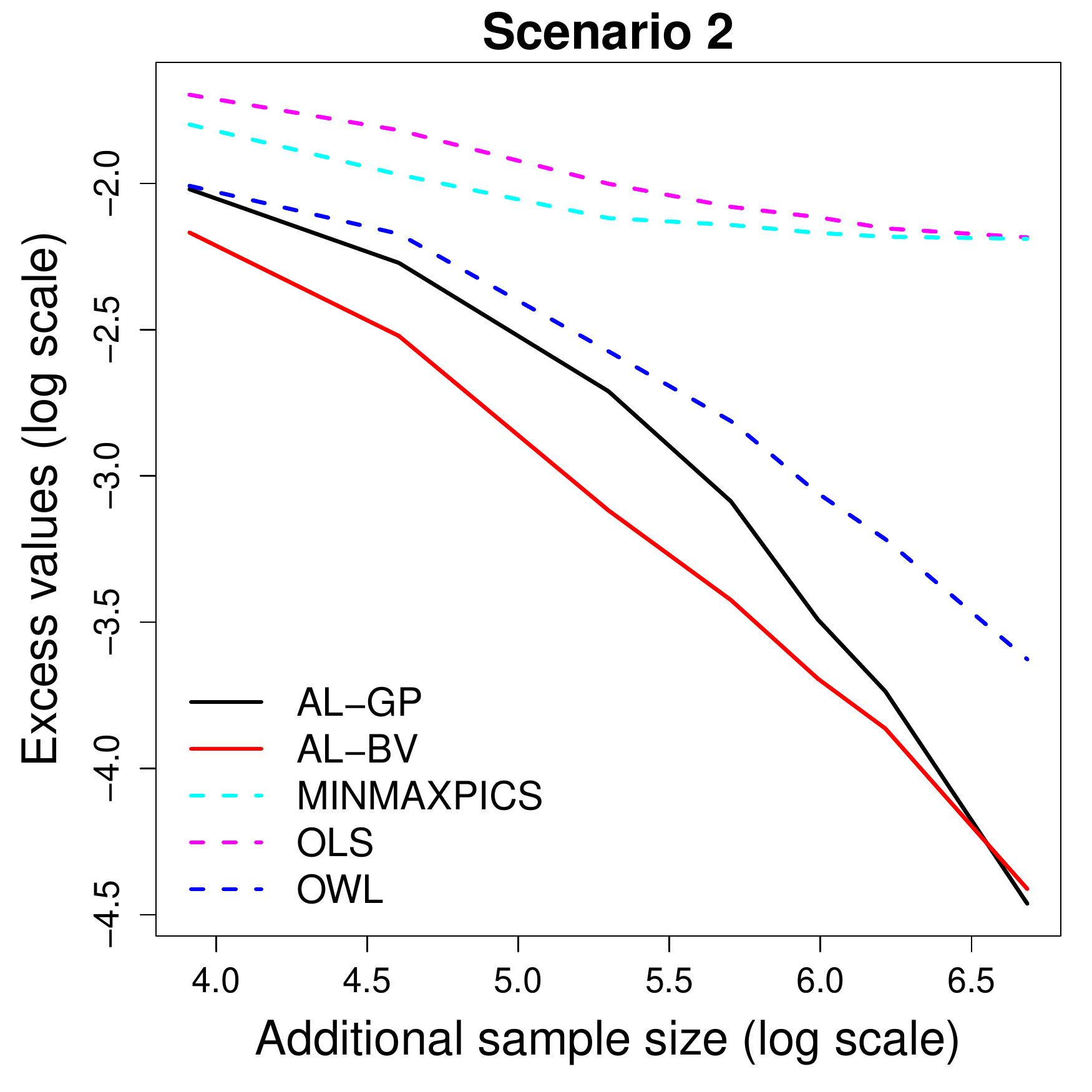}
\includegraphics[width=3.2in,height=3.2in]{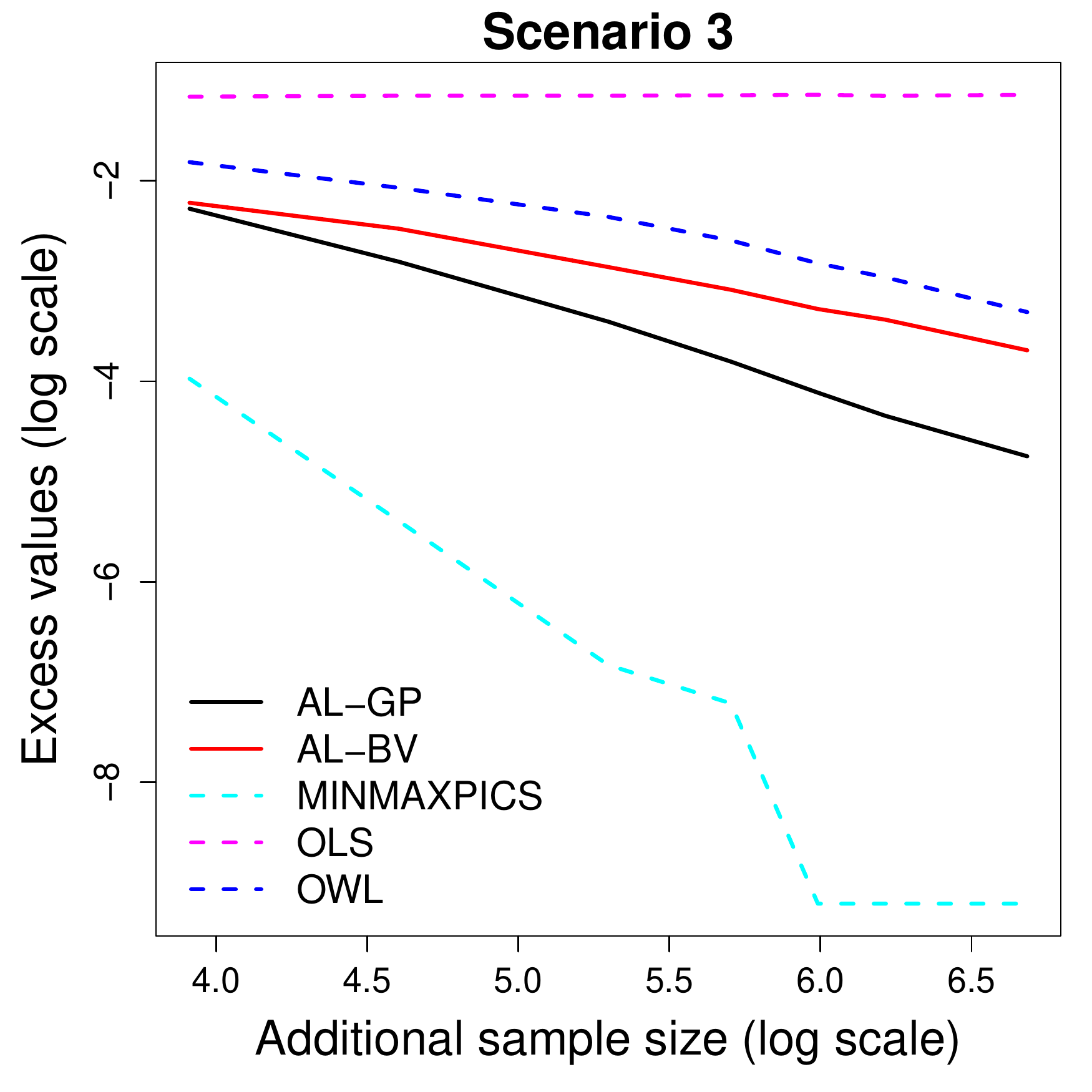}
\includegraphics[width=3.2in,height=3.2in]{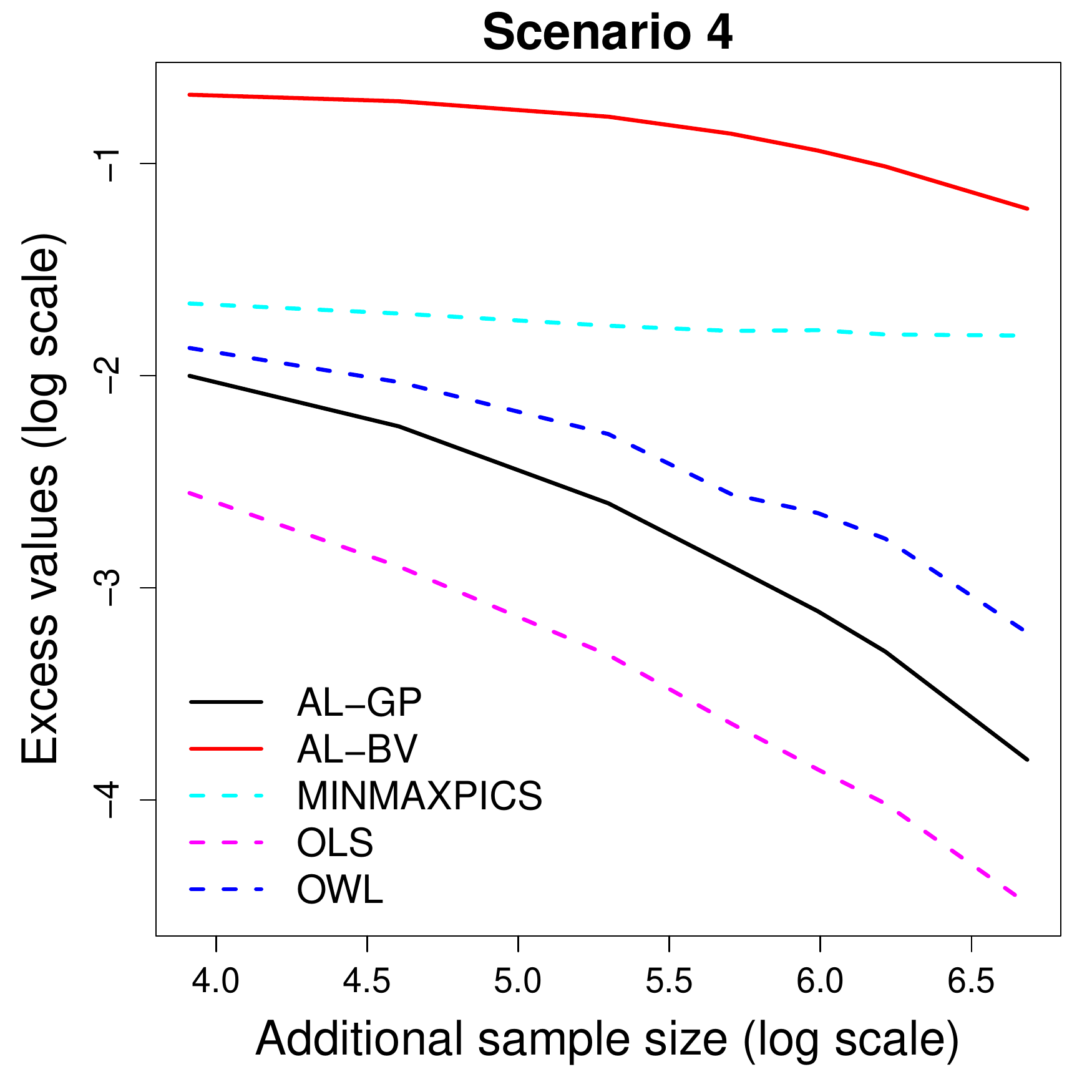}
\end{center}
\end{figure}

\section{Real Data Analysis} \label{scn_data}

\subsection{Nefazodone--CBASP Clinical Trial}

We apply the proposed active learning methods to analyze the data from the Nefazodone--CBASP clinical trial \citep{keller:CBASP2000}. The randomized trial was conducted to compare the efficacy of three treatments for  nonpsychotic chronic major depressive disorder (MDD), namely Nefazodone, cognitive
behavioral-analysis system of psychotherapy (CBASP), or a combination of Nefazodone and CBASP. CBASP requires twice-weekly on-site visits to the clinic, and thus imposes a significant burden on patients compared with Nefazodone alone. Hence, we compare Nefazodone with the combination treatment, and investigate whether CBASP is necessary for all patients. We perform a complete case analysis. The score on the 24-item Hamilton rating scale for depression (HRSD) was the primary outcome, with higher scores indicating more severe depression. Data from 441 patients were available, with 218 patients randomized to Nefazodone and 223 to the combined treatment group. We consider three covariates for tailoring the treatment: the baseline HRSD scores, the alcohol dependence, and the HAMA somatic anxiety scores. The latter two covariates were selected by \cite{gunter2007variable} as important variables for optimal treatment decision making.

All the patients are used for the OLS and OWL analysis.  To mimic an active clinical trial, we first randomly select 50 patients as the initial dataset. We then sequentially examine the remaining patients, and drop those who do not satisfy the selection criteria.  The recruitment stops once an additional 300 patients are enrolled, or all the patients have been examined. The eligible patients are included in the dataset, assuming that they have been randomized  in an active clinical trial. The treatments they are randomized to in this hypothetical active clinical trial are the actual treatments they received in the completed trial. In particular, 289 and 350 patients were used to construct the optimal ITRs using AL-BV and AL-GP respectively. The estimated ITRs from the different methods were applied to the whole data set to calculate the average HRSD scores due to the ITRs, with smaller values being preferable. Here, the ``average HRSD scores due to the ITRs'' are defined to be
 $
 {\pn[\{\widetilde R I(A=D(X)\}/\pi(A)]}/{\pn[\{I(A=D(X)\}/\pi(A)]},
 $
where $\widetilde R$ denotes the HRSD scores and $\pi(A)$ is the probability of being assigned treatment $A$. AL-BV recommends combination therapy for 299 patients, giving an average HRSD score of 8.86.  AL-GP recommends this therapy for 386 patients, giving an average HRSD score of 9.72. OLS and OWL recommend this therapy for all the patients, yielding a higher score of 10.89. Hence, the treatment rules produced by the active clinical trials not only lead to a higher overall benefit, but also reduce the time and monetary commitments for the patients.

\begin{figure}[h!]
\caption{Mean cross-validated HRSD scores (the lower the better) against additional sample sizes. The initial sample size was set to 50.}
\label{fig_mdd}
\begin{center}
\includegraphics[width=3.5in,height=3.5in]{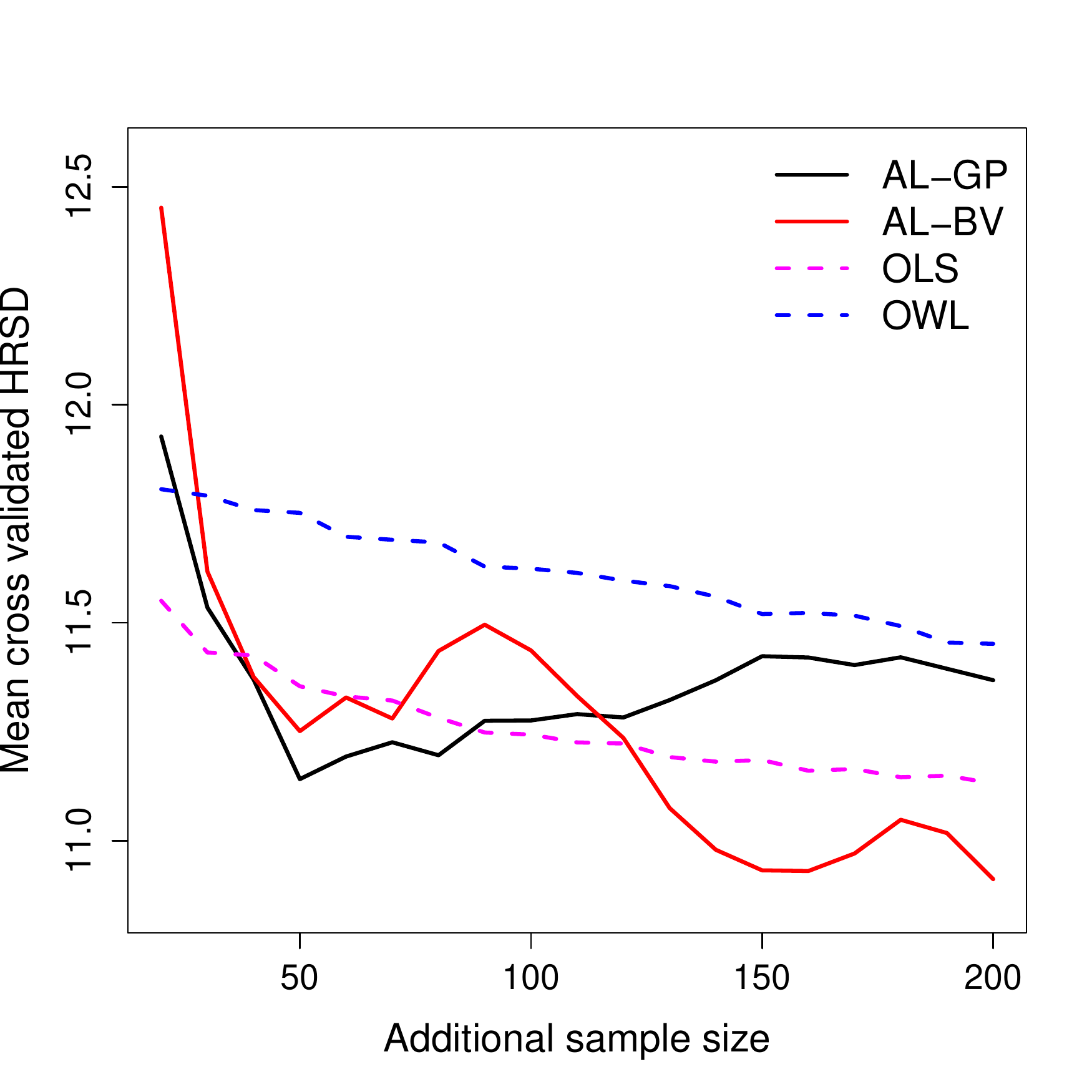}
\end{center}
\end{figure}

We then used a five-fold cross-validation analysis to avoid potential overfitting. The data set was partitioned into five subsets. Four of the five subsets were used as training data to construct the optimal ITR, and the remaining subset was used as the validation set to evaluate the estimated rule. In the training subset, we applied both active learning and passive learning methods, i.e., OLS and OWL, to construct the optimal ITRs. The initial sample size was set to 50 for the active learning methods. The number of additional recruited patients was $20, 30,\ldots,200$; these were adaptively selected from the rest of the training samples. For the passive learning methods, we randomly chose $70, 80,\ldots,250$ patients from the training data and conducted the estimation.
The process was repeated 200 times, and we recorded the average cross-validated values for each sample size. The results are presented in Figure \ref{fig_mdd}. The active learning methods initially lead to higher HRSD scores, but they catch up and continue to improve as the sample size grows. In particular, we can see that the HRSD scores from our methods decrease faster than those of the other methods do. After the additional sample size reaches 120, with the total sample size at 170, the ITR identified by AL-BV yields the lowest score, and this value can be further improved with larger sample sizes.

\subsection{Twelve-step Intervention on Stimulant Drug Use}
These data come from a randomized clinical trial that aims to evaluate the effectiveness of 8-week group intervention plus individual 12-step facilitative intervention for reducing stimulant drug use \citep{donovan2013stimulant}. Individuals with stimulant-use disorders were randomly assigned to treatment as usual (TAU) or to TAU integrated with Stimulant Abuser Groups to Engage in 12-step (STAGE-12) intervention.

The primary outcome of interest is the number of days of self-reported stimulant drug use over the three- to six-month post-randomization period,  where a smaller value is preferable. We use seven baseline variables to evaluate the patients and to construct the optimal ITR: age, average number of days per month of self-reported stimulant drug use in the three months prior to randomization, baseline alcohol use, drug use, employment status, medical status, and psychiatric status composite scores on the addiction severity index (ASI), where the ASI composite score, ranging from 0 (no endorsement of any problems) to 1 (maximal endorsement of all problems), is usually taken as an indication of problem severity; it is perceived to guide the treatment decision  \citep{mcgahan1986addiction}.

\begin{figure}[h!]
\caption{Mean cross-validated outcomes (number of  days of self-reported stimulant drug use over the three- to six-month post-randomization period, the lower the better) against additional sample sizes. The initial sample size was set to 50.}
\label{fig_stimulant}
\begin{center}
\includegraphics[width=3.5in,height=3.5in]{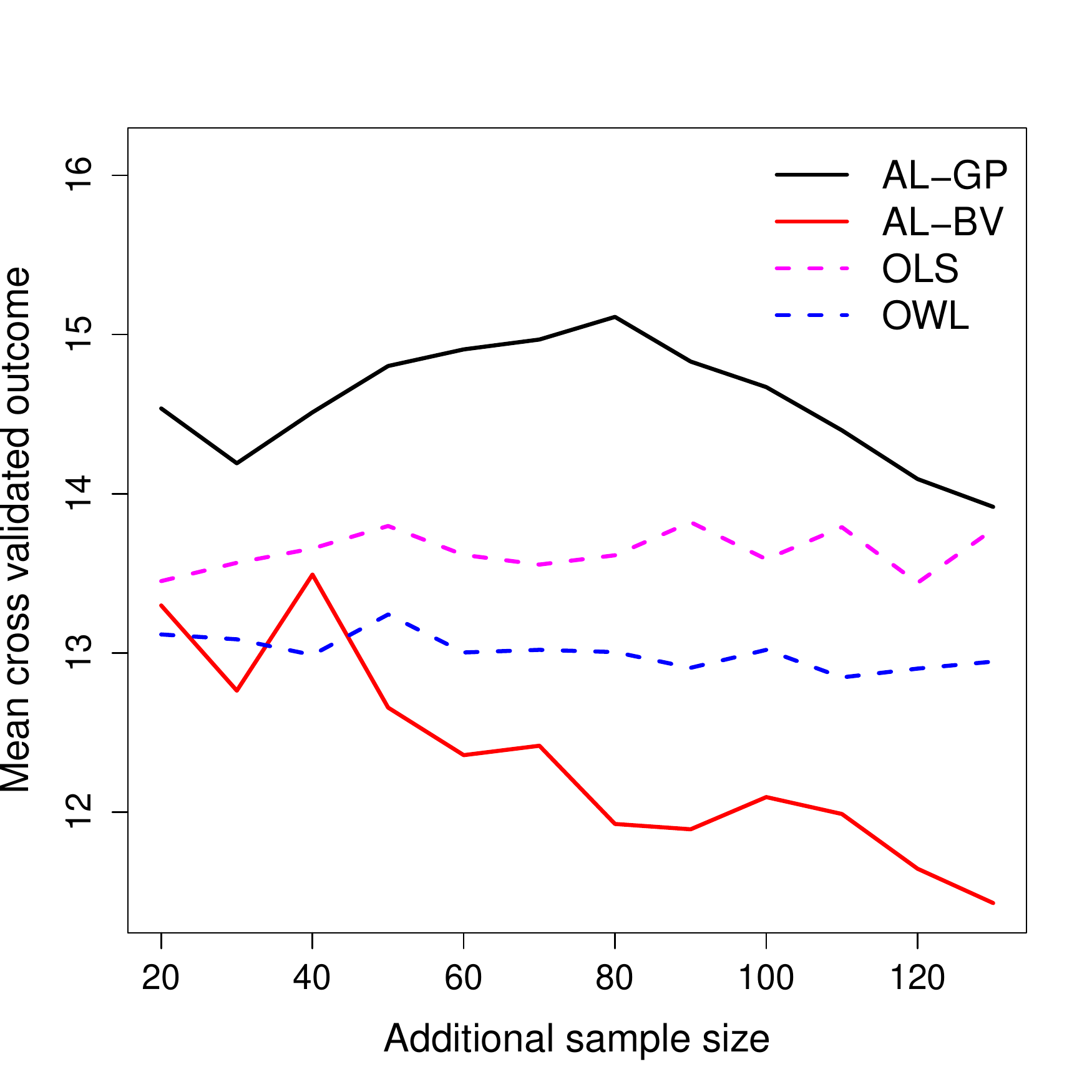}
\end{center}
\end{figure}

After removing the missing data, we have 305 participants in total. We evaluated the different methods on the whole data set, with the initial sample size set to 50 and the additional sample size set to 200. AL-BV and AL-GP assigned 123 and 90 patients to the STAGE-12 group, with expected outcomes of 11.3 and 12 respectively. OLS and OWL gave overall averages of 11.8 and 12.7 days. We also calculated the cross-validated number of days of drug use over the three- to six-month post-randomization period. Since the outcome was count data, with considerable zero-inflation, overdispersion, and a nonlinear trend, we anticipated that the active learning method using a bias-variance tradeoff for the estimation would lead to the best results. Indeed, as shown in Figure \ref{fig_stimulant}, AL-BV outperforms the other methods with a fast overall decreasing trend. While the Gaussian assumption is severely violated in this example, AL-GP still improves with sample size. We note that OLS and OWL do not improve noticeably as the sample size grows.

\section{Sample Size Consideration}
In order to design a future study, it is important to plan ahead a sample size, which is sufficient to guarantee that we will obtain an ITR very close to the optimal one using the proposed active clinical trial.  Here, we provide a data-driven method that gives a preliminary assessment of the required sample size. Further work is warranted to derive the sample size formula for such an exploratory trial.  Let $V_0$ be an average patient outcome based on the standard care. Our goal is to determine a minimal sample size $N^*$ such that: (1) the power to reject the hypothesis $V(\D^*)=V_0$ is at least $1-\beta$ when $V(\D^*)\geq (1+\rho)V_0$; and (2) $Pr(|V(\hat \D_{N^*})-V(\D^*)|\leq \epsilon) \geq 1-\alpha$ \citep{laber2015sizing}, where $ \epsilon$, $\alpha$, and $\rho$ are prespecified constants and $\hat \D_N$ is the resulting estimator  based on $N$ samples from the active clinical trial. Since we have constructed a non-asymptotic error bound for $|V(\hat \D_{N})-V(\D^*)|$ in terms of the sample size $N$ in Theorem \ref{thm:conv}, we suggest to obtain $N^*$ by inverting the above error bound such that $Pr(|V(\hat \D_{N^*})-V(\D^*)|\leq \min(\rho V_0, \epsilon)) \geq 1-\alpha.$ By doing so, we are able to provide finite sample evidence that the optimal individualized treatment rule yields a larger average benefit than the standard care, and, moreover, the average benefit of the estimated rule is approximately optimal. The constant $\gamma \in [0, d]$ will be determined based on prior knowledge, or can be set to a range of values to examine the resulting sample sizes.  The unknown constant $\widetilde C$ can be determined via simulations. Given $d, \gamma, \theta$ and $\widetilde C$, the required sample size $N^*$, such that with probability greater than $1-\alpha$, $|V(\hat \D_{N^*})-V(\D^*)|\leq \min(\rho V_0, \epsilon)$, can be found by setting $ \widetilde C N^{*-\frac{1+\gamma}{2+d-\gamma}}\l(\log(N^*/\alpha)\r)^{\theta}=\min(\rho V_0, \epsilon)$. The required sample sizes vary by the allocated initial sample size $N_0$. Therefore, our sample size calculation is conditional on a specified $N_0$. We will illustrate the proposed strategy by designing a future study to explore the optimal ITRs for patients with MDD. The Nefazodone-CBASP clinical trial is used as the basis for planning.

Consider two treatment options, Nefazodone only versus the combined treatment. The average HRSD score in the combined treatment group is 10.96. It is desirable to develop an ITR that can at least reduce the expected HRSD score that would be achieved by 15\% with probability greater than 80\%, i.e., $\rho V_0\approx 1.7$ and $\alpha=0.2$. We will use three tailoring variables with $d=3$.. To determine $\tilde C$, we adopt a bootstrap method. We will bootstrap the data, a total of 441 patients, 10000 times. An active clinical trial, with 50 initial and 100 additional patients, will be implemented with each bootstrapped sample to estimate the optimal ITR. The average HRSD score under each estimated ITR will be recorded. We denote these scores as $\hat V_1^B, \ldots, \hat V_{10000}^B$. Let $k^*=\argmin_{k=1,\ldots, 10000} \hat V_k^B$. $\hat V_{k^*}^B$ will be used as an estimate of $V(\D^*)$. Denote the 80 percentile of $|\hat V_k^B -\hat V_{k^*}^B|, k\neq k^*, k=1,\ldots, 10000$ as $\widehat{Diff}$. $\widetilde C$ can be obtained using $\tilde C = N^\frac{1+\gamma}{2+d-\gamma}\l(\log(N/\alpha)\r)^{-\theta} \times \widehat{Diff}$, where $N=150$ and $\alpha=0.2$. We will assess different $\gamma$'s with $\gamma=0.5, 1, 1.5, 2, 2.5$ and 3, which lead to different $\theta$'s and different estimates of $\widetilde C$. Using the outlined strategy, we suggest to plan the sample size for the future study according to Table \ref{tb_samplesize}.
\begin{table}[h]
\begin{center}
\begin{tabular}{cccccc}
\hline\hline
 $\epsilon$ & $\gamma=0.5$ & $\gamma=1$ & $\gamma=1.5$ & $\gamma=2$ & $\gamma=2.5$  \\
 \hline
1.7 &   165 & 160 &  155 &  155 &  150  \\
1.6 &   215  & 190 &  175 &  165 &  160  \\
1.5 &   280  & 225 &  200 &  180 &  170  \\
\hline\hline
\end{tabular}
\caption{Planned sample size for different combinations of $(\gamma,\epsilon)$.}
\label{tb_samplesize}
\end{center}
\end{table}

\section{Discussion}
\label{scn_discussion}

In this paper, we propose an active clinical trial with the goal of constructing favorable ITRs at a minimal cost. This new paradigm is distinct from the standard clinical trial framework that is designed for treatment evaluation.  Along with the new designs, we also present new analysis and inference tools that are often practically useful and theoretically efficient. Two methods are presented to construct the confidence interval in the algorithm. The kernel smoothing method is better in the situation where  a few important tailoring variables are known a priori. However, due to the curse of dimensionality, it cannot handle a covariate space of moderate to high dimension. The Gaussian process regression approach performs better in this situation. However, the kernel smoothing method does not assume a specific underlying distribution of the outcomes, whereas the Gaussian process regression approach requires the outcome to follow a Gaussian distribution, which may not be appropriate for discrete outcomes. Since the goal of the proposed trial is to explore the optimal treatment rules, it is a learning stage. The results should be further validated in a confirmatory trial.

In practice, we may also want to include individuals who would be known to  benefit from the treatment. Instead of dropping certain patients whose confidence interval for the contrast function does not contain 0, we can enroll all patients with different priority (quantified by probability). Specifically, we can prioritize those patients, whose confidence interval of the contrast function contains 0, by assigning larger probabilities, while enrolling patients, whose confidence interval for the contrast function does not contain 0, with lower probabilities. Under this more flexible framework, individuals who benefit from the treatment are also included in the trial such that the marginal treatment effects can be investigated. In addition, when there is a small difference between two medications, treatment choice will likely depend on other considerations such as cost, side effects and etc. Our framework can take into account those factors that can be quantified. By allowing a ratio of the cost per treatment relative to the cost of a worse disease outcome, say $\delta$, which reflects the patients' tolerance of treatment burden relative to tolerance of the disease burden \citep{vickers2007method,huang2014personalized}, we maximize $V(D)-\delta P(A=1)$. The optimal rule is instead given by $D^*(x)=\sign\{f^*(x)-\delta\}$. In this case, the active clinical trial will enroll patients from whom the value of $f^*(x)$ is close to $\delta$, since these patients are close to the decision boundary and thus most informative.

We next discuss two directions for future research. First, our method can easily be extended to incorporate multi-category treatments, where patients are recruited based on the minimum differential treatment effect of all the pairwise comparisons. For continuous treatments, one possibility is to discretize the continuous treatment into different percentiles and then apply the extended method for multi-category treatments. Second, in this paper, we assume that the outcomes of the previous patients have been observed before we enroll the next patient. In fact, the proposed active clinical trials can be conducted with delayed responses, provided the required estimates (for the contrast function) can be updated as data become available, i.e., some responses can be collected during the study period. We can also update the estimates after groups of responses instead of individual responses. There have been some recent developments on handling the problem of delayed outcomes in phase I--II trials, for example, by treating them as missing values and applying imputation strategies \citep{jin2014using}. We would like to explore such options in the future. There are two other interesting extensions that we are pursuing: (i) a more general contrast function that can accommodate a high-dimensional or discrete covariate; (ii) a dynamic treatment regime for a sequence of treatment rules \citep{Murphy03optimaldynamic, Robins04optimalstructural}.

\appendix

\section{Technical Proofs}

\subsection{Intrinsic dimension of $\supp(\Pi)$}
\label{asm:a3}

We explain the meaning of ``intrinsic dimension'' introduced in Assumption~(A3) here.
We say that $\supp(\Pi)$ possesses a \textit{tree decomposition} $\m T=\l\{T_{i,j}, \ i\geq 1, \ j=1\ldots J(i)\r\}$ if
\begin{enumerate}
\item $T_{1,1}=\supp(\Pi)$, and $\{T_{i,j}\}_{j=1}^{J(i)}$ forms a disjoint partition of $\supp(\Pi)$  for all $i\geq 1$.
\item Nested partition: $\forall i\geq 2, j=1,\ldots, J(i)$, there exists a unique $1\leq k\leq J(i-1)$ such that $T_{i,j}\subset T_{i-1,k}$;
\item Bounded diameter: for all $i\geq 1, 1\le\ j\leq J(i)$,
\[
{\rm diam}(T_{i,j}):=\sup_{x,y\in T_{i,j}}\|x-y\|_2\leq K_1 2^{-i}
\]
for some $K_1=K_1(\Pi)$.
\item Regularity: for any $i\geq 1, 1\le j\leq J(i)$ and $0<r\leq 2^{-i}$ the following holds: there exists a $1\leq d\leq p$ ($d$ is the intrinsic dimension) such that for all $x\in T_{i,j}$,
 \begin{eqnarray}
 c_1 r^d\leq\Pi(B(x,r)\cap T_{i,j})\leq c_2 r^d\label{ass:int}
 \end{eqnarray}
for some $0<c_1(\Pi)\leq c_2(\Pi)<\infty$ which are independent of $i,j$.
Here, $B(x,r)$ is the Euclidean ball of radius $r$ centered at $x$.
\end{enumerate}

A simple example that gives a good intuition to the tree decomposition is the uniform distribution over the unit cube in $\mb R^p$.
In this case, the tree decomposition is given by partitioning the unit cube into dyadic cubes and $d=p$.
If $\supp(\Pi)$ is contained in a proper subspace $W$ of $\mb R^p$, then $d\leq \dim(W)$.

Let $\m B_i$ be the sigma-algebra generated by the collection of sets $\{T_{i,j}, \ j=1,\ldots, J(i)\}$ (a partition of $\supp(\Pi)$ on the level $i$). A regular approximation of ${\rm AS}_k$ in Algorithm \ref{alg1} is given by
$\act_k:=\bigcap\left\{A: \ A\in \m B_{m_{k-1}}, \ A\supset {\rm AS}_k\right\}$.



\subsection{Properties of Kernel Estimate}
\subsubsection{Preliminaries}\label{sec:pre}

For a measurable set $S\subset\supp(\Pi)$, define $\Pi_S(dx):=\Pi(dx|x\in S)$ as the conditional distribution on $S$, and set
\begin{align}
\nonumber
\mb Q_{h}(x|S):=\int\limits_{\mb R^p} K_h\l({x-y}\r)d\Pi_S(y).
\end{align}
Since $\Pi$ is assumed to be known, we can directly compute $Q_h(x|S)$ now. Accordingly, we modify the original kernel estimate for $\eta_j$, i.e., (\ref{fhat}), as follows: let $\{(X^{(i)},A^{(i)},R^{(i)}), \ i=1\ldots N\}$ be an i.i.d. sample from the conditional joint distribution of $(X,A,R)$ given that $X\in S$, and set
\begin{align}
\label{eq:estim}
&\widehat\eta_{j}(x;h,S)=\frac{1}{N}\sum\limits_{i=1}^N \frac{R^{(i)} I\{A^{(i)}=j\} K_h\l(x-X^{(i)}\r)}{Q_h(x|S)P(A^{(i)}=j)}, \ j=\pm1, \\
&\nonumber
\widehat f(x;h,S)=\widehat\eta_{1}(x;h,S)-\widehat\eta_{-1}(x;h,S).
\end{align}
We will discuss properties of these estimators in below.

Let $h>0$, $S\in \m B_j$ and $h\leq 2^{-j}$, and define
\begin{eqnarray}
\mb Q_{h,m}(x|S):=\int\limits_{\mb R^p} \|x-y\|_2^{m} K_h\l({x-y}\r)d\Pi_S(y)\label{def:q}
\end{eqnarray}
We next study the upper and lower bounds of $\mb Q_{h,m}(x|S)$ based on Assumptions (A1)-(A4). Since $K$ is bounded and compactly supported, there exists $R=R_K>0$ such that $K(x)\leq \|K\|_\infty I\{x\in B(0,R_K)\}$. Let $F>0$ be a large enough constant, namely, $F^d\geq 2c_2/c_1$. Recall that $c_1, c_2$ are defined in (\ref{ass:int}). Note that Assumption (A2) implies the following:
\begin{align}
\nonumber
\mb Q_{h,m}(x|S)&\geq \ell_K\int\limits_{B(x,h)\cap S}\|x-y\|_2^m d\Pi_{S}(y) \\
&\nonumber \geq
\ell_K\l(h/F\r)^m \l(\int\limits_{B(x,h)\cap S} d\Pi_{S}(y)-\int\limits_{B(x,h/F)\cap S} d\Pi_{S}(y)\r)\\
&\label{eq:app10}
\geq
\ell_K\l(h/F\r)^m \frac{\l(c_1 h^d-c_2(h/F)^d\r)}{\Pi(S)}\geq
\frac{1}{2F^m}\ell_K c_1\frac{ h^{d+m}}{\Pi(S)}:=c_3 \frac{h^{d+m}}{\Pi(S)},
\end{align}
and
\begin{align}
\label{eq:app11}
\mb Q_{h,m}(x|S)\leq & \|K\|_\infty \int\limits_{B(x,R_Kh)\cap A}\|x-y\|_2^m d\Pi_{S}(y)\leq \|K\|_\infty R_K^{m+d} c_2 \frac{h^{d+m}}{\Pi(S)}:=c_4 \frac{h^{d+m}}{\Pi(S)}.
\end{align}

In what follows, we will set $Q_h(x|S):=Q_{h,0}(x)$ for brevity.

\subsubsection{Some bounds for the kernel estimators}\label{app:ker}

In this subsection, we derive basic concentration inequalities for the kernel estimators of $\eta_j(x)=\mb E[R|A=j,X=x]$, $j=\pm1$ restricted to $S$, i.e., $\widehat\eta_{j}(x;h,S)$ defined in (\ref{eq:estim}). The proof of the results can be found in supplementary materials.
\begin{lemma}
\label{lem:sup-norm}
For all $t>0$ satisfying $t+d^2\log(1/h)\leq nh^d$, with probability $\geq 1-2e^{-t}$,
\[
\sup\limits_{x\in \supp(\Pi)\cap S}|\widehat\eta_j(x; h)-\eta_j(x)|\leq C\left(h + \sqrt{\frac{\Pi(S)(t+d^2\log(1/h))}{nh^d}}\right),
\]
 where $C=C(M,c_1,c_2,L,L_K, \|K\|_{\infty},\ell_K,R_K)$ is a constant.
\end{lemma}

The following Corollary is immediate:

\begin{corollary}
\label{cor:1}
Set $h_n:=\left\{\Pi(S)(t+d\log(n/\Pi(S)))/n\right\}^{1/(d+2)}$.
Then, under assumptions of Lemma \ref{lem:sup-norm}, with probability $\geq 1-4e^{-t}$,
\[
\sup\limits_{x\in \supp(\Pi)\cap S}|\widehat f(x;h_n)-f^\ast(x)|\leq 4C h_n,
\]
where constant $C$ is the same as in Lemma \ref{lem:sup-norm}.
\end{corollary}

\subsection{Proof of Theorem \ref{thm:conv}}

\subsubsection{Comparison inequality}

Our Lemma~\ref{lem:comparison} below illustrates the connection between the risk $V(\widehat D)-V(D^\ast)$ of a treatment rule $\widehat D(x)=\sign(\widehat f(x))$ and the sup-norm $\|\widehat f-f^\ast\|_{\infty,\supp(\Pi)}$.
\begin{lemma}
\label{lem:comparison}
Under the margin assumption (A4),
\begin{align}
&
\nonumber
V(\widehat D)-V(D^\ast)\leq C(\gamma)\|(\widehat f-f^\ast)\m I\left\{\sign(\widehat f)\ne \sign(f^\ast)\right\}\|_{\infty,\supp(\Pi)}^{1+\gamma}.
\end{align}
\begin{proof}
It is easy to see that
$V(\widehat D)-V(D^\ast)=2\mb E\l(|f^\ast(X)| I\{\widehat D(X)\ne D^\ast(X)\}\r)$.
The rest of the argument repeats Lemma 5.1 in \citet{Audibert2007Fast-learning-r00}.
\end{proof}
\end{lemma}

\subsubsection{Main proof}
Our main goal is to control the size of the set $\act_k$ defined by Algorithm \ref{alg1}.
In turn, these bounds depend on the size of the confidence bands for $f^\ast(x)$ (denoted by $\delta_k$).
Suppose $L\leq N$ is the number of iterations performed by the algorithm before termination.

Let $N_k^{\act}:=\lfloor N_k\cdot \Pi(\act_k)\rfloor$
be the number of labels requested on the $k$-th iteration of the algorithm. We first claim that the following bounds hold uniformly for all $1\leq k\leq L$ with probability at least
$
1-\alpha
$:
\begin{align}
\label{eq:Z1}
& \nonumber
\|f^\ast-\widehat f_k\|_{\infty,\act_k}\leq C_1\left(\frac{\log(N/\alpha)+d\log(N_k)}{N_k}\right)^{1/(d+2)}, \\
&
\Pi(\act_{k})\leq C_2
\left(\frac{\log(N/\alpha)+d\log(N_{k-1})}{N_{k-1}}\right)^{\gamma/(d+2)},
\end{align}
where $C_j=C_j(M,c_1,c_2,L,L_K, \|K\|_{\infty},\ell_K,R_K,\gamma), \ j=1,2$. This claim will be proved later.

Let $\m E$ be the event of probability $\geq 1-\alpha$ on which both inequalities of (\ref{eq:Z1}) hold, and assume that it occurs.
Second inequality of (\ref{eq:Z1}) implies, together with the fact that $N_k=2N_{k-1}$ by definition, that the number of randomized subjects on each step
$1\leq k\leq L$ satisfies
\[
N_k^{\act}=\lfloor N_k\Pi(\act_k)\rfloor\leq 2 N_{k-1}^{\frac{2+d-\gamma}{2+d}}
\left(\log(N/\alpha)+d\log(N_{k-1})\right)^{\gamma/(d+2)}
\]
with probability $\geq 1-\alpha$.
If $N$ is the maximum number of randomized subjects the algorithm is allowed to request, then
\begin{align*}
N\leq&\sum\limits_{k=0}^L N_k^{\act}\leq 2\left(\log(N/\alpha)+d\log(N_{L})\right)^{\gamma/(d+2)}
\sum_{k=0}^L N_{k}^{\frac{2+d-\gamma}{2+d}}
\leq \\
&
C_3(\gamma,d)\left(\log(N/\alpha)+d\log(N_{L})\right)^{\gamma/(d+2)}N_{L}^{\frac{2+d-\gamma}{2+d}},
\end{align*}
and one easily deduces that on the last iteration $L$ we have
\begin{equation}
\label{Z2}
N_L\geq c(\gamma,\Pi,d)\left(\frac{N}{\log(N/\alpha)}\right)^{\frac{2+d}{2+d-\gamma}}.
\end{equation}
Recall that $N_L$ is defined in Algorithm~\ref{alg1}.

To obtain the risk bound of the theorem from (\ref{Z2}), we apply
Lemma \ref{lem:comparison}:
\begin{equation}\label{final}
V(\widehat D)-V(D^\ast)\leq
C(\gamma)\left\|(\widehat f_L-f^\ast)\cdot  I\left\{\sign(\widehat f_L)\ne D^\ast\right\}\right\|_{\infty,\supp(\Pi)}^{1+\gamma}.
\end{equation}
Since $\left\{\sign(\widehat f_L)\ne D^\ast\right\}\subseteq \act_L$ whenever bounds (\ref{eq:Z1}) hold,
it remains to estimate $\|\widehat f_L-f^\ast\|_{\infty,\act_L}$.
Recalling the first inequality of (\ref{eq:Z1}) once again (for $k=L$), we get
\[
\|(\widehat f_L-f^\ast)\|_{\infty,\act_L} \leq
C_1\left(\frac{\log(N/\alpha)+d\log(N_L)}{N_L}\right)^{1/(d+2)}\leq \widetilde C N^{-\frac{1}{2+d-\gamma}}\l(\log(N/\alpha)\r)^{q},
\]
where $q=\frac{4+2d-\gamma}{(2+d)(2+d-\gamma)}$, which together with (\ref{final}) yields the final result.

It remains to show both inequalities of (\ref{eq:Z1}).
We start with the bound on $\|\widehat f_k-f^\ast\|_{\infty, \act_k}$.
First, note that by construction, for every $k\geq 1$ the samples $(X^{(i,k)},A^{(i,k)},R^{(i,k)}), \ i=1\ldots \lfloor N_k\Pi(\act_k)\rfloor$ are conditionally independent given the data $\bigcup_{i=1}^{k-1}S_i$ collected on steps $1,\ldots,k-1$, with conditional distribution of $X^{(i,k)}$ being $\Pi_{\act_k}$.
Thus we can apply Corollary \ref{cor:1} conditionally on $\bigcup_{i=1}^{k-1}S_i$ with $t=\log\frac{4N}{\alpha}$ to get that with probability $\geq 1-\alpha/N$,
\[
\|\widehat f_k-f^\ast\|_{\infty, \act_k}\leq 4C\left(\frac{\log\frac{\alpha}{4N}+d\log(\lfloor N_k\Pi(\act_k)\rfloor/\Pi(\act_k))}{\lfloor N_k\Pi(\act_k)\rfloor/\Pi(\act_k)}\right)^{1/(d+2)}\leq  8C h_k.
\]
It remains to integrate the bound with respect to the distribution of $\bigcup_{i=1}^{k-1}S_i$:
\begin{align*}
P\left(\|\widehat f_k-f^\ast\|_{\infty, \act_k}\geq 8Ch_k\right)=\mb E P\left(\|\widehat f_k-f^\ast\|_{\infty, \act_k}\geq 8Ch_k\,\big|\,\bigcup_{i=1}^{k-1}S_i\right)\leq
\frac{\alpha}{N}.
\end{align*}
The union bound over all $1\leq k\leq L\leq N$ gives the result.

Finally, we will prove the second inequality of (\ref{eq:Z1}), the bound for the size of the active sets $\act_k$.
This is the place where assumption (A3) on the tree decomposition and margin assumption (A4) play the key role.
To obtain the bound, we will compare two estimators of $f^\ast$: the first is the kernel estimator $\widehat f_k$ constructed by the Algorithm \ref{alg1} on step
$k$, and the second is the piecewise-constant estimator $\bar f_k$ with similar approximation properties to $\widehat f_k$.
Namely, $\bar f_k$ is the $L_2(\Pi)$ - projection of $f^\ast$ on the linear space of piecewise-constant functions of the form
$g(x)=\sum\limits_{j=1}^{J(m_k)}\alpha_j I\{T_{m_k,j}\}(x), \ \alpha_j\in \mb R$. Recall that $T_{i,j}$ is defined in the tree decomposition of Section~\ref{asm:a3}. As a result, we will be able to relate the ``active sets" associated to these estimators, taking advantage of the fact that the active set associated to $\bar f_k$ is always a union of the sets from a collection $\{T_{m_k,j}, \ j=1\ldots J(m_k)\}$.

Let $\m E_1$ be the event of probability $\geq 1-\alpha$ on which
$\|\widehat f_k-f^\ast\|_{\infty,\act_k}\leq \delta_k$ for any $k\geq 0$, where $\delta_k=4Ch_k$.
Assume that $\m E_1$ occurs.

The following inclusions hold (for the definition of $AS_{k+1}$, see Algorithm \ref{alg1}):
\begin{align}
\label{eq:z3}
&
\l\{x: |f^\ast(x)|< \delta_k/2\r\}\subseteq AS_{k+1}\subseteq \l\{x: |f^\ast(x)|< 5\delta_k/2\r\}.
\end{align}
Indeed,
\[
|f^\ast(x)|< \delta_k/2\implies |\widehat f_k(x)|< \delta_k/2+|f^\ast(x)-\widehat f_k(x)|< \frac{3}{2}\delta_k\implies x\in AS_{k+1}
\]
and
\[
x\in AS_{k+1}\implies |\widehat f_k(x)|< \frac{3}{2}\delta_k\implies |f^\ast(x)|< \frac{5}{2}\delta_k.
\]
For all $x\in T_{m_k,j}$, set $\bar f_k(x):=\frac{1}{\Pi(T_{m_k,j})}\int\limits_{T_{m_k,j}}f^\ast(y)d\Pi(y)$, and note that
\begin{align*}
|f^\ast(x)-\bar f_k(x)|\leq &\frac{1}{\Pi(T_{m_k,j})}\int\limits_{T_{m_k,j}}|f^\ast(y)-f^\ast(x)|d\Pi(y)\leq
\frac{2L}{\Pi(T_{m_k,j})}\int\limits_{T_{m_k,j}}|x-y|d\Pi(y)\leq \\
&2L\,{\rm diam}(T_{m_k,j})\leq 2LK_1 2^{-m_k}\leq 4LK_1 h_k,
\end{align*}
where the last two inequalities follow from part 3 of assumption (A3) given in Appendix~\ref{asm:a3}, and from the definition of $m_k$.
Define $\tau_k:=\max\l(5\delta_k,4LK_1 h_k\r)\leq C_5\delta_k$,
$
\bar {\m F}_{k+1}:=\{f: \ |f(x)-\bar f_k(x)|\leq (3/2)\tau_k, \  \forall x\in \act_k\}
$
to be the band of size $(3/2)\tau_k$ around $\bar f_k$, and
\[
\bar A_{k+1}:=\left\{x\in \act_k: \ \exists f_1,f_2\in \bar {\m F}_{k+1}, \sign(f_1(x))\ne\sign(f_2(x))\right\}.
\]

By a reasoning similar to above, we have the inclusions
\begin{align}
\label{eq:z4}
\l\{x: |f^\ast(x)|< \tau_k/2\r\}\subseteq \bar A_{k+1}\subseteq \l\{x: |f^\ast(x)|< 5\tau_k/2\r\}.
\end{align}
Moreover,  by the definition of $\tau_k$, we have the inequality $5\delta_k/2\leq \tau_k/2$. Hence (\ref{eq:z3},\ref{eq:z4}) imply that
$AS_{k+1}\subseteq \bar A_{k+1}$.
It remains to note that
\begin{enumerate}
\item $\bar A_{k+1}$ is the union of the sets from a collection $\{T_{m_k,j}, \ j=1\ldots J(m_k)\}$, hence $\bar A_{k+1}\supseteq \act_{k+1}$;
\item By (\ref{eq:z4}) and assumption (A4),
\[
\Pi(\act_{k+1})\leq \Pi(\bar A_{k+1})\leq \Pi(\l\{x: |f^\ast(x)|< 5\tau_k/2\r\})\leq K_2(5\tau_k/2)^\gamma\leq C_6 \delta_k^\gamma,
\]
hence proving the claim.

\end{enumerate}


\newpage
\section*{Supplementary Materials for ``Active Clinical Trials for Personalized Medicine"}
\section*{S.1 Additional simulation results}

\subsection*{S.1.1 Simulation setups with normal biomarkers}
In this section, we first present two additional examples where biomarkers have normal distributions (Scenarios 5 and 6). Particularly,  $X_{1}$ and $X_{2}$ are generated according to $N(0,1)$. Treatments $A_1$, $A_2$ are randomly generated from $\{-1,1\}$ with equal probability 0.5.  Outcomes $R$ are generated as follows.
\begin{enumerate}
\item[5.] $R\sim N(1+2X_1+0.5(X_1^2-0.25)A,1);$
\item[6.] $R\sim N(1+2X_1^2+X_2+2(\log(|X_2|)+\sqrt{|X_1|}-1)A,1).$
\end{enumerate}

The results are presented in Figure \ref{sim_fig}. The outcome models are fairly complex in both scenarios. Hence the results from OLS are unsatisfactory. We prespecify 4 subgroups, with $X_1$ and $X_2$ dichotomized at 0, to implement the approach proposed in \citet{deng2011activedeveloping}.  Since these pre-specified subgroups are incorrect to inform the optimal treatment rule, the ITRs estimated by their approach are not consistent for the optimal. Both active learning methods show competitive performances in both scenarios. Kernel approach performs slightly worse than the Gaussian process regression. This phenomenon is anticipated since the bounded support assumption required in theoretical guarantee is violated.

\begin{figure}[h!]
\caption{Excess values (log scale). Initial size was set at 50.}
\label{sim_fig}
\begin{center}
\includegraphics[width=3.2in,height=3.2in]{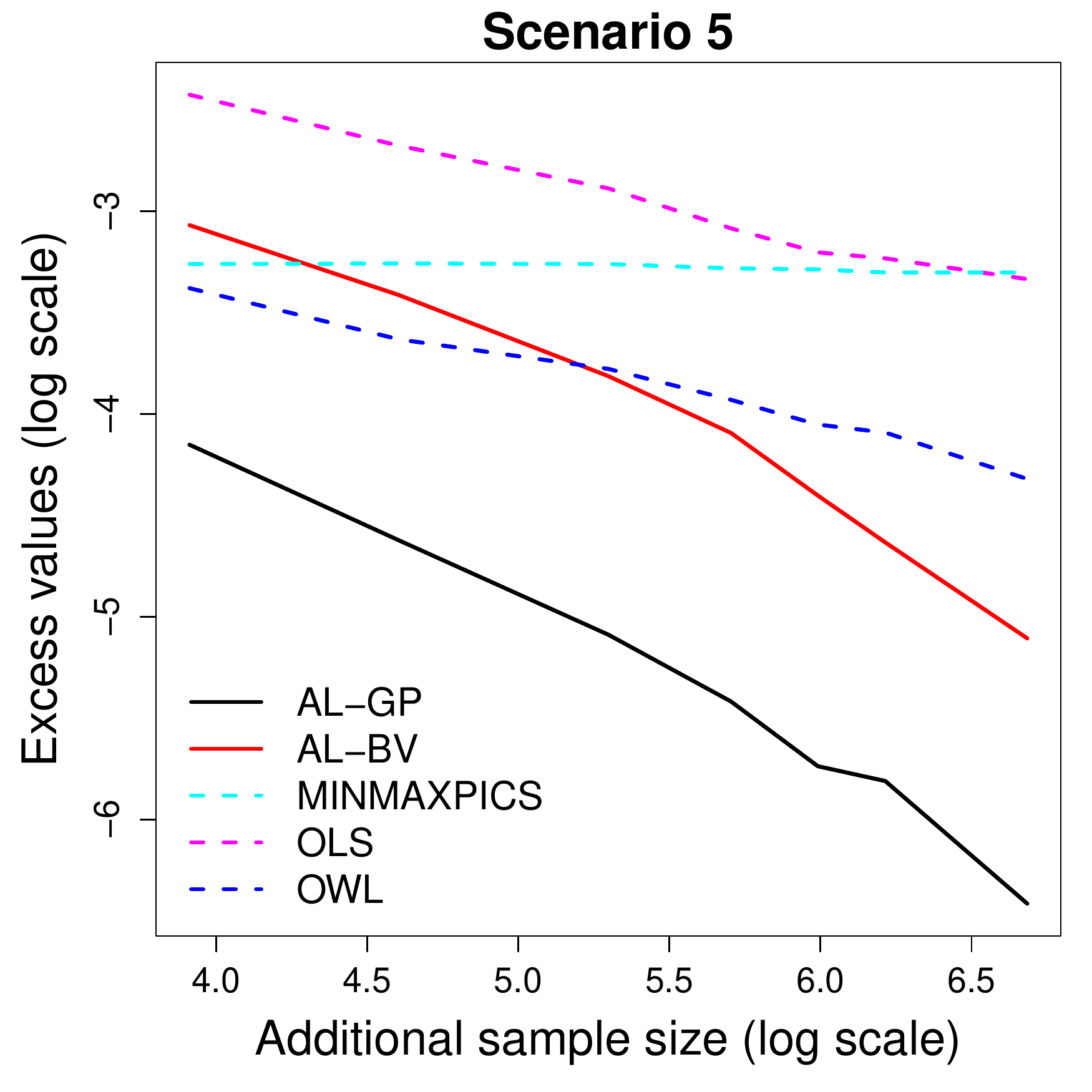}
\includegraphics[width=3.2in,height=3.2in]{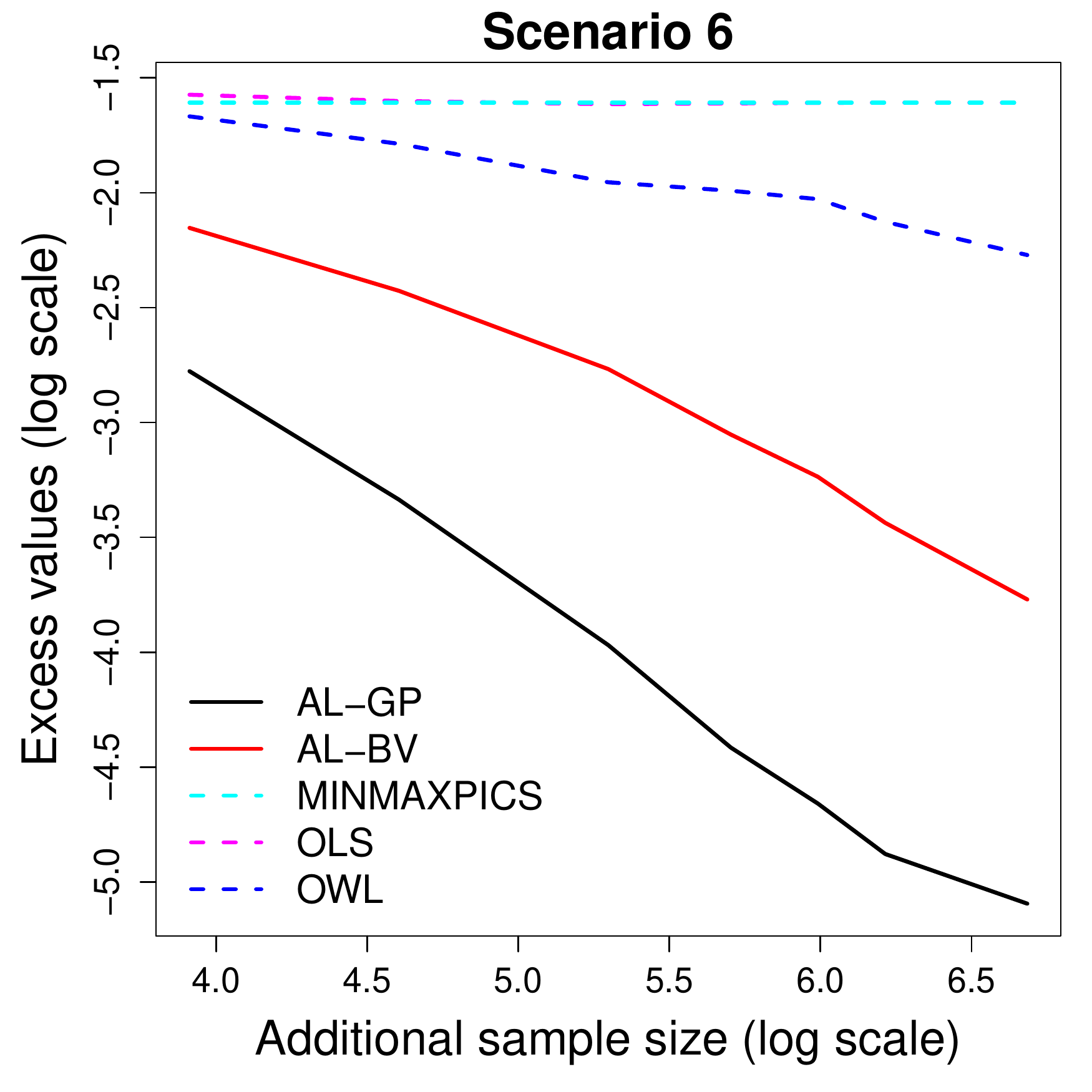}
\end{center}
\end{figure}

\subsection*{S.1.2 Sensitivity analysis }
\begin{figure}[h!]
\caption{Excess values (log scale) with different initial sizes in Gaussian process regression method. `N0 flexible' means $N_0= 2\lfloor\sqrt{n}\rfloor$.}
\label{sim_in}
\begin{center}
\includegraphics[width=3.2in,height=3.2in]{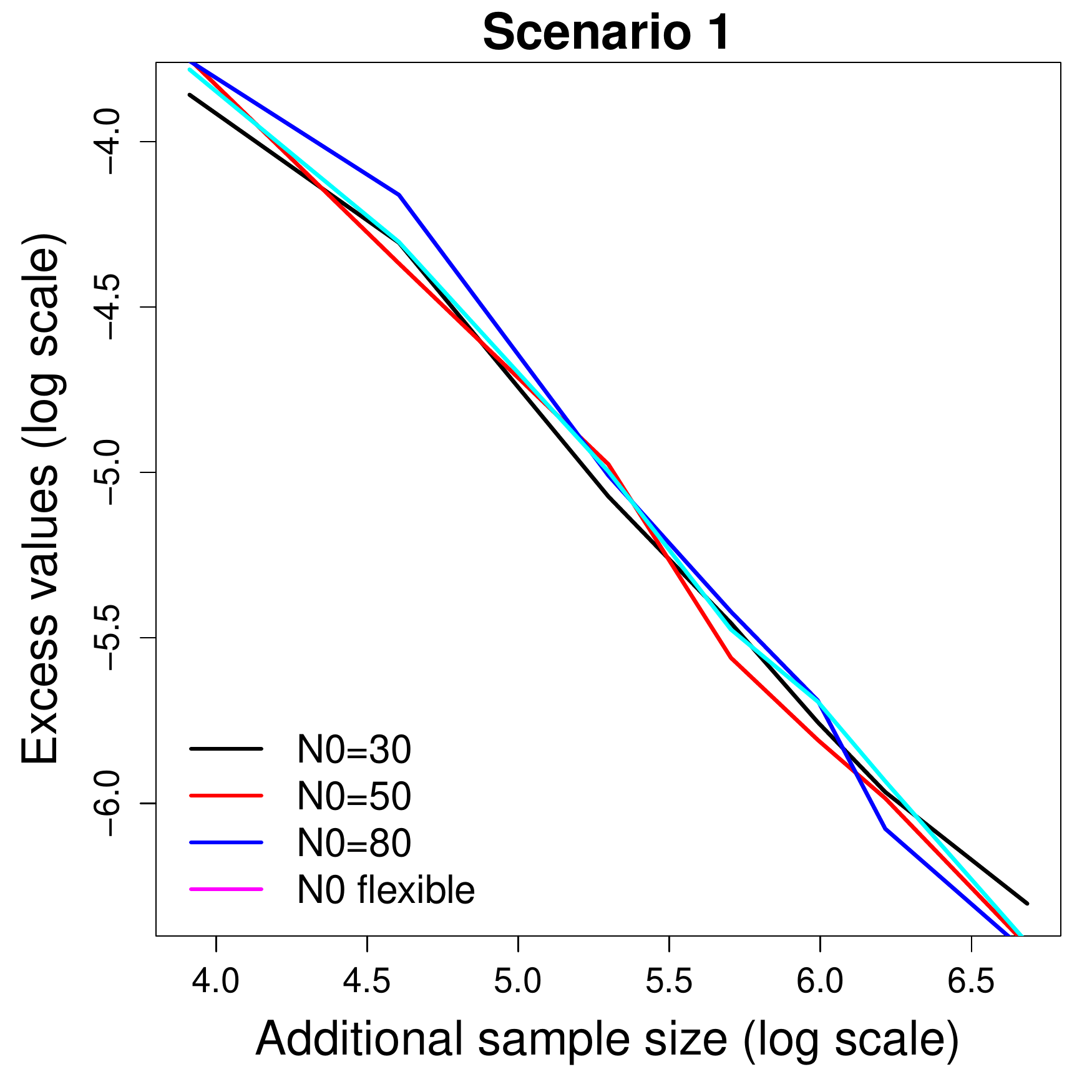}
\includegraphics[width=3.2in,height=3.2in]{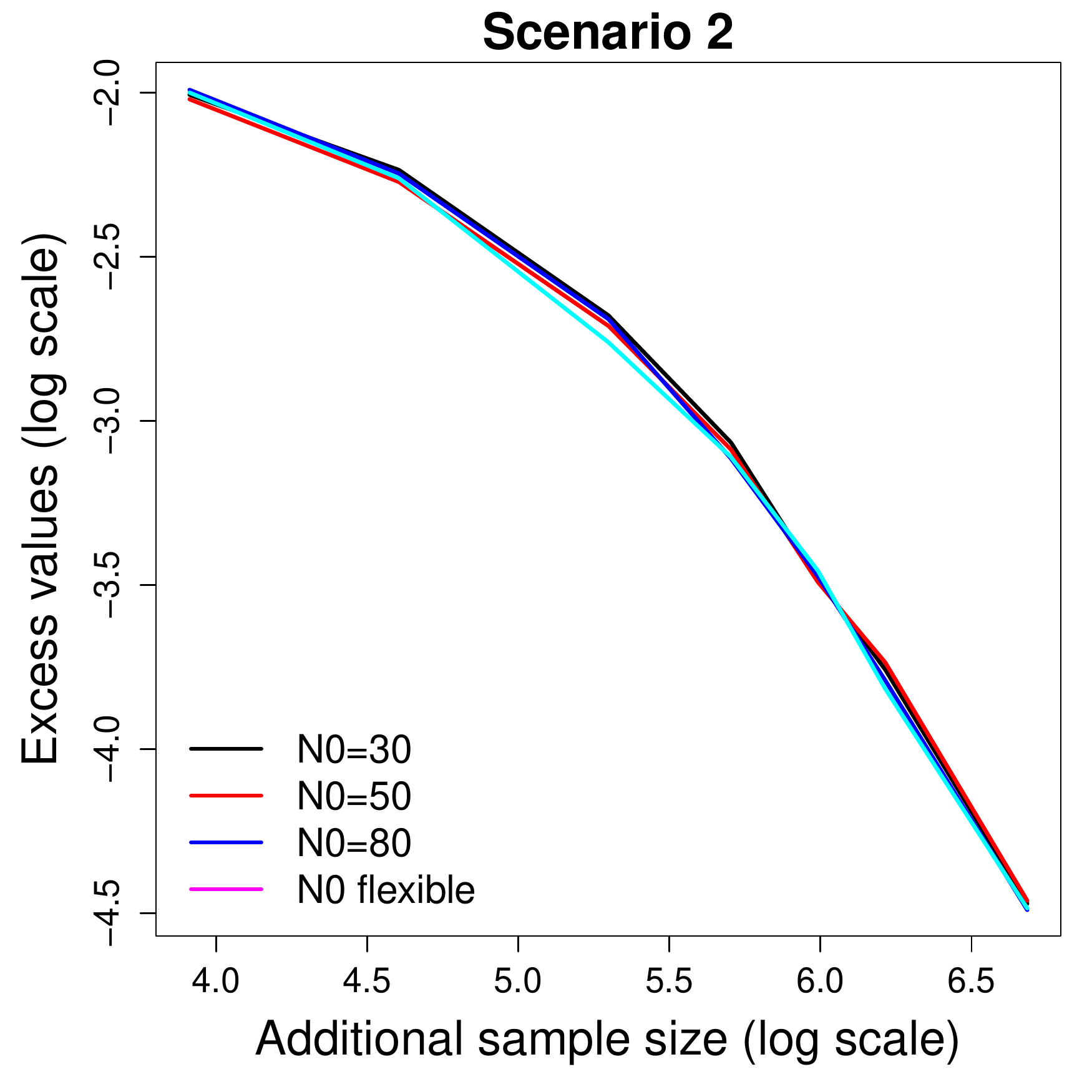}
\includegraphics[width=3.2in,height=3.2in]{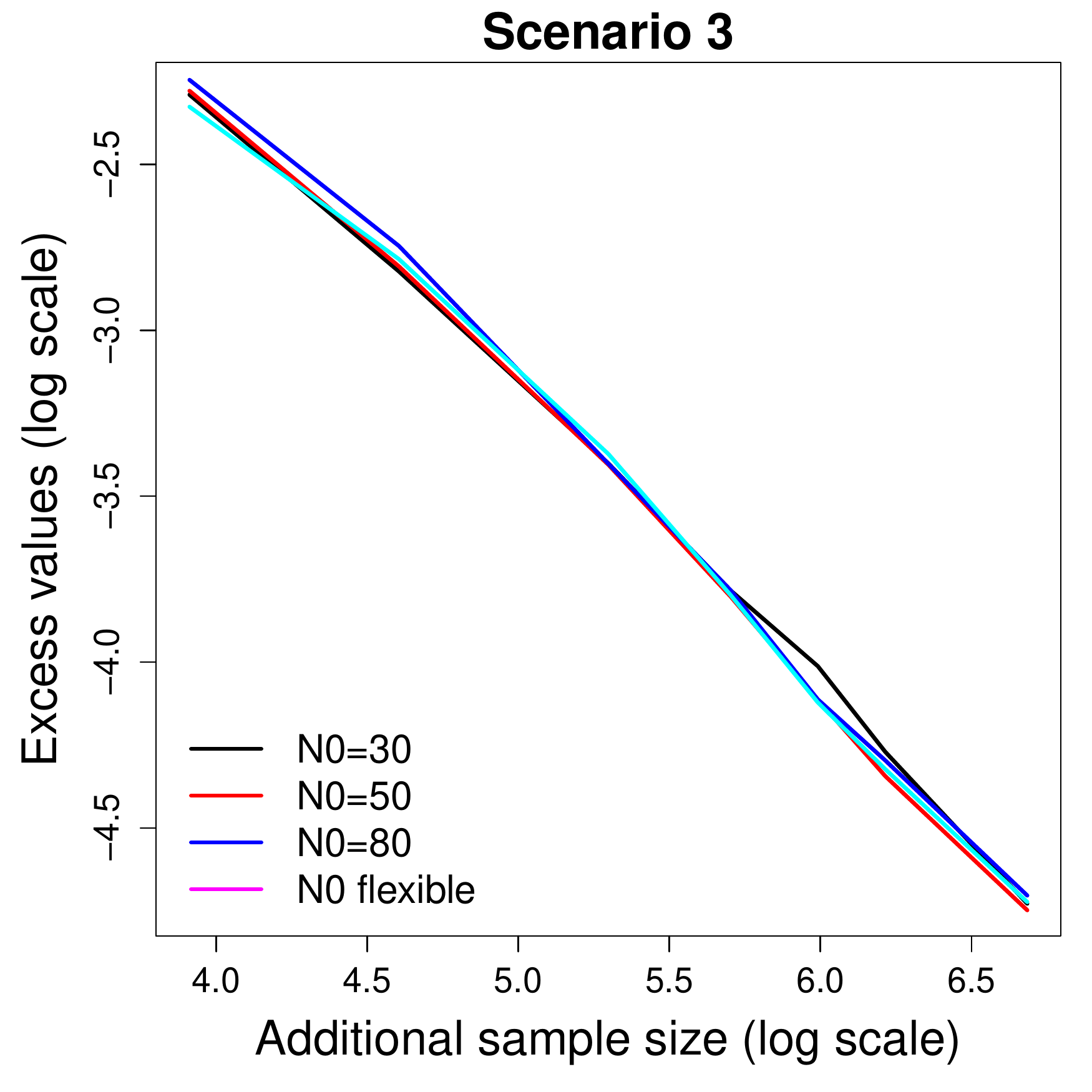}
\includegraphics[width=3.2in,height=3.2in]{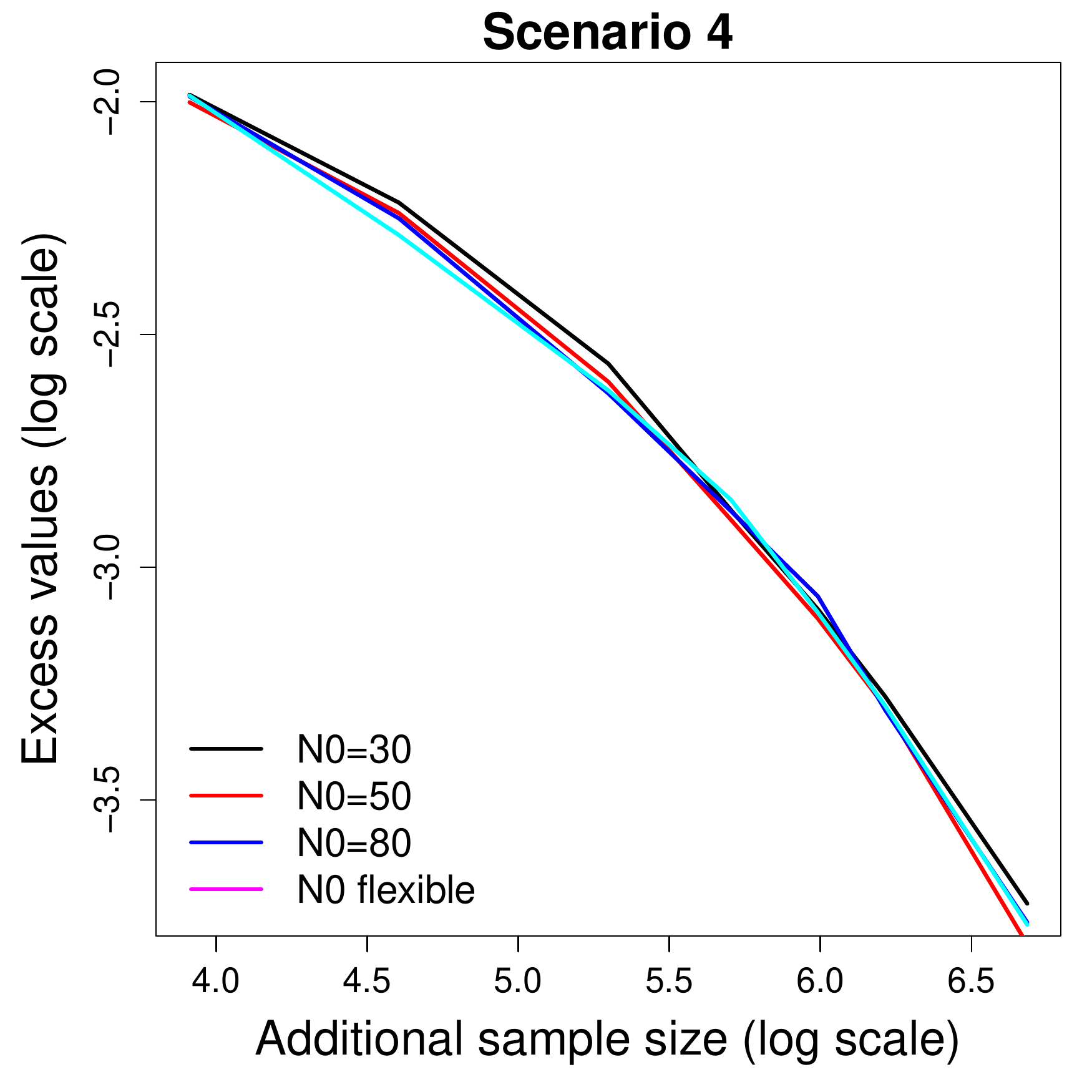}
\end{center}
\end{figure}

In this section, we conduct a thorough sensitivity analysis to investigate whether the performances of the proposed methods depend on the choices of initial sample size $N_0$. Basically, any number that is ``significantly smaller'' than the total allocated budget should lead to the same theoretical asymptotic behavior. Indeed, we vary $N_0$ from 30, 50 to 80 and find that the results are insensitive to the choice. We present the sensitivity analyses for Gaussian process regression method in Figure \ref{sim_in}. The results for kernel approach are similar (data not shown).

\begin{figure}[h!]
\caption{Excess values (log scale) with different confidence parameters in kernel method. Initial size was set at 50. }
\label{sim_t}
\begin{center}
\includegraphics[width=3.2in,height=3.2in]{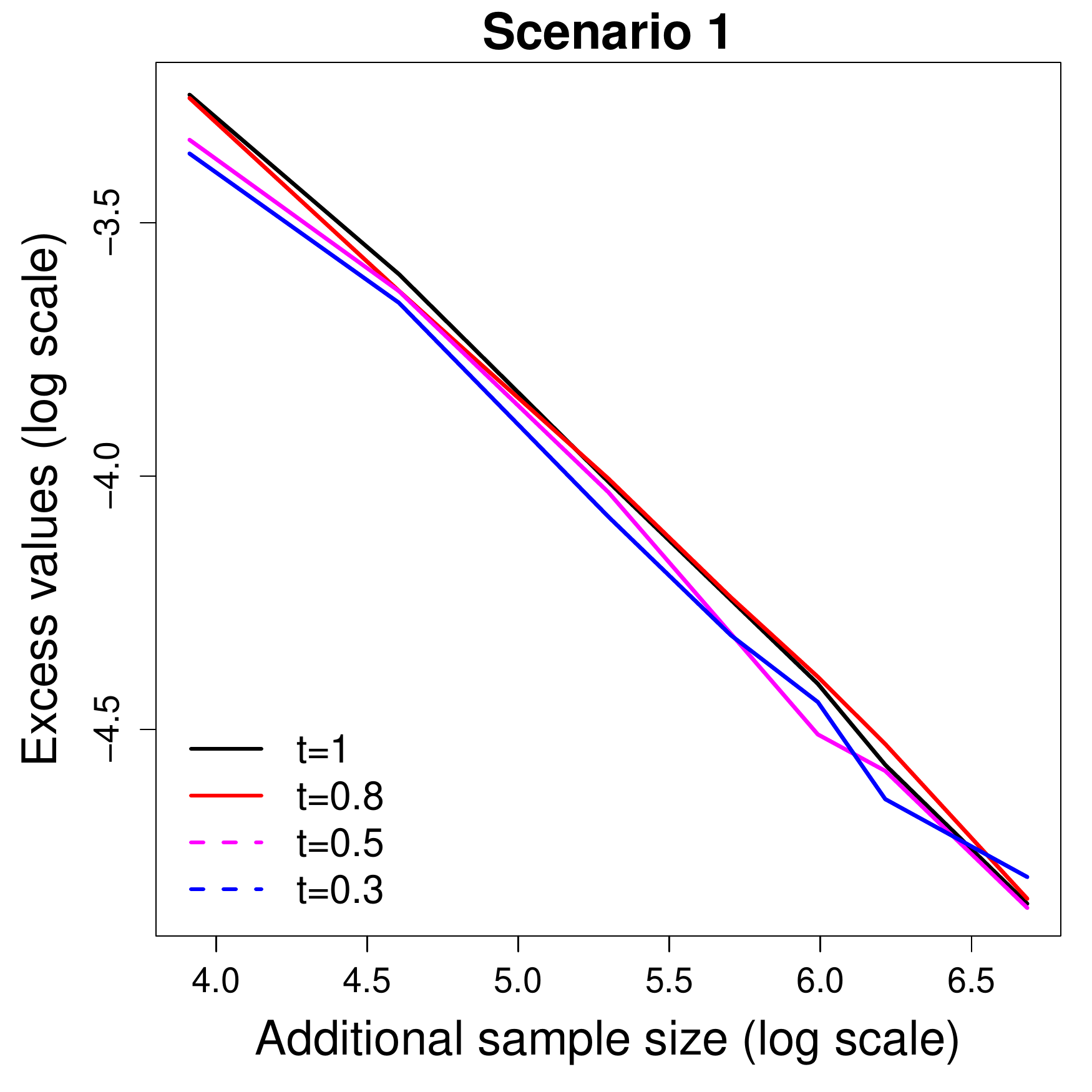}
\includegraphics[width=3.2in,height=3.2in]{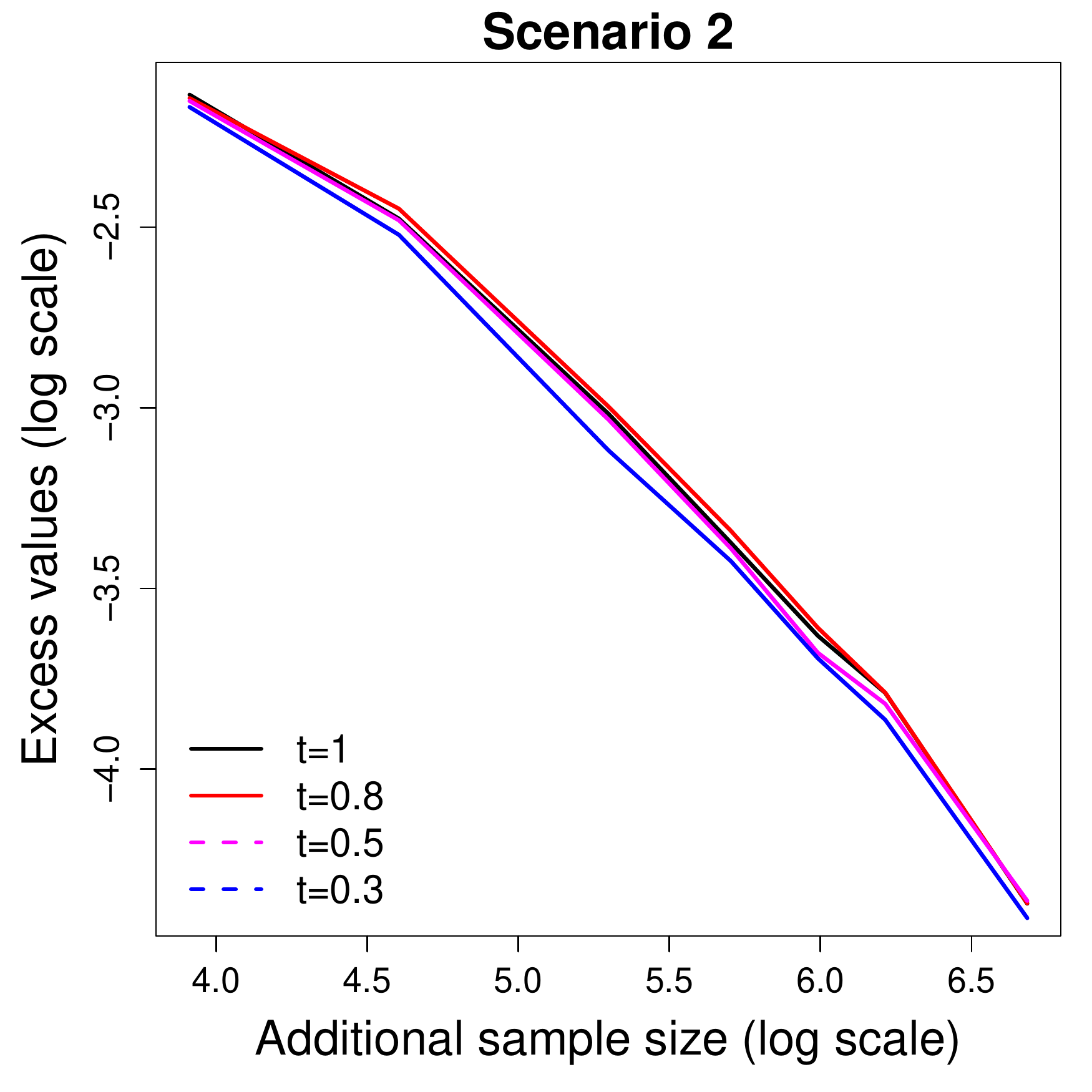}
\includegraphics[width=3.2in,height=3.2in]{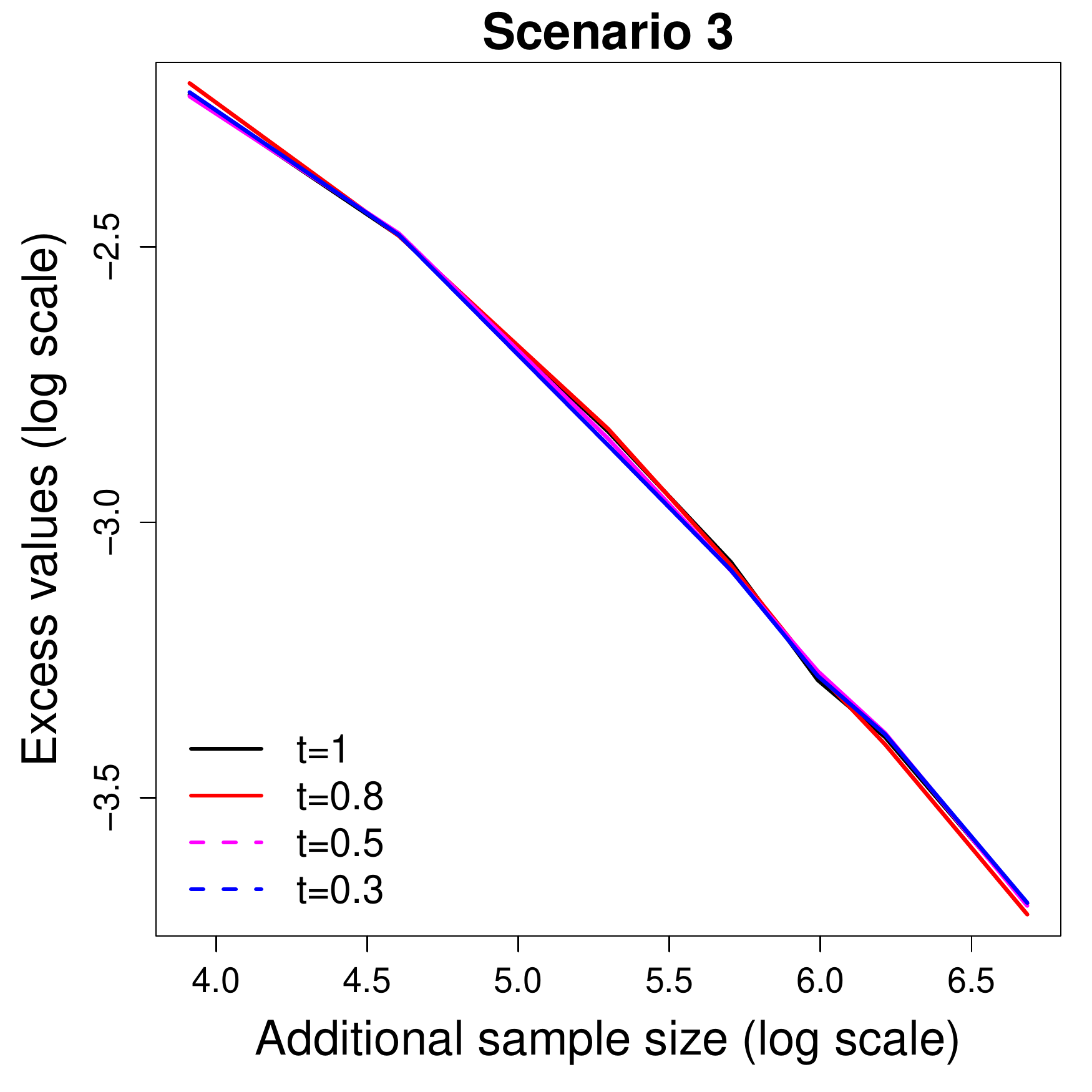}
\includegraphics[width=3.2in,height=3.2in]{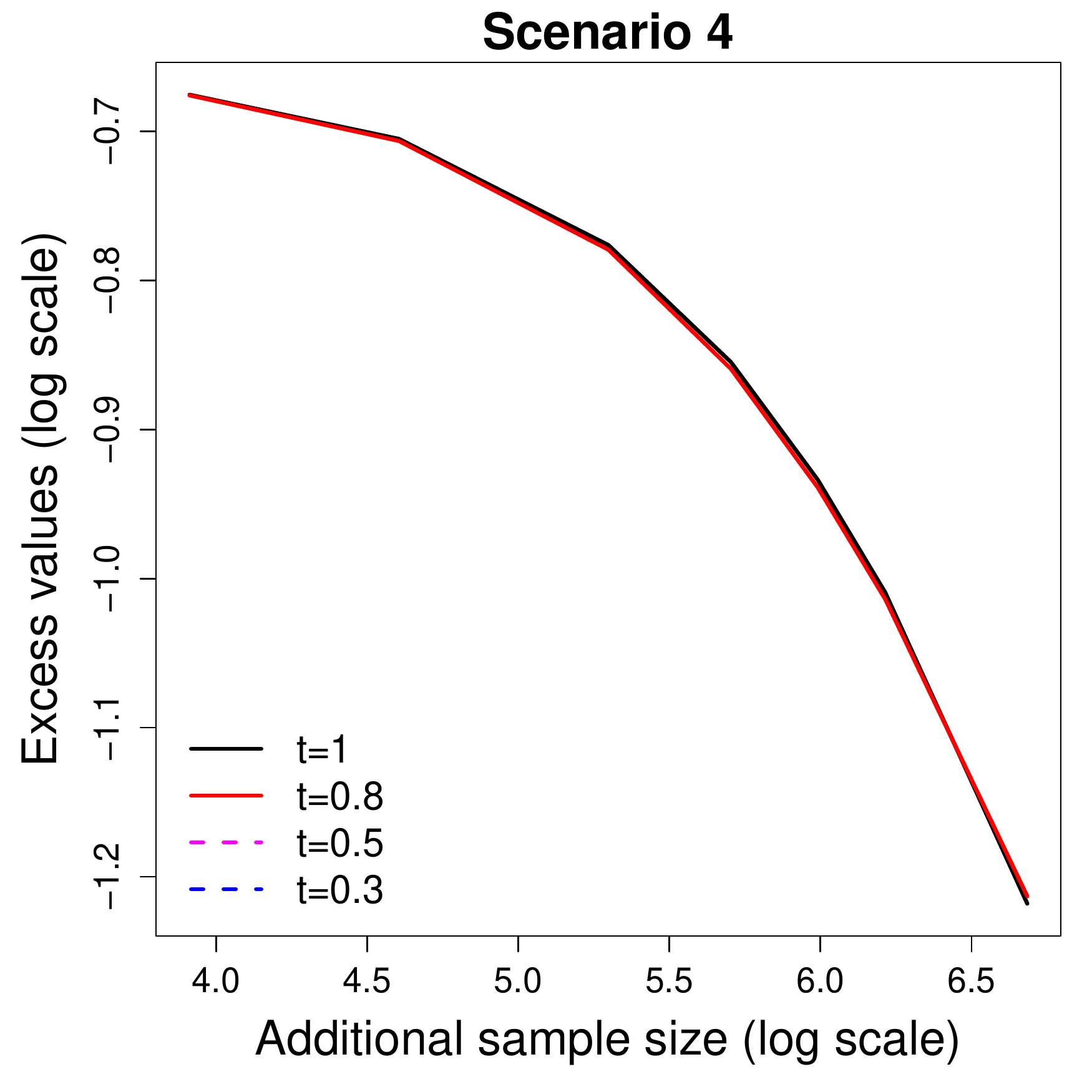}
\end{center}
\end{figure}

We also investigate how the confidence level $\delta(x_0)$ in the step of Algorithm 1 affects the performance of the algorithm. The confidence level $\delta(x_0)$ is controlled by the parameter $t$, where $\delta(x_0)=tLh_n(x_0)$.  We vary $t$ over the set $\{0.3, 0.5, 0.8, 1\}$, and implicitly assume $L=1$. The results presented in Figure \ref{sim_t} clearly show that algorithm performances are robust to the choice of the confidence parameter.

\subsection*{S.1.3 Doubly-robust estimator}
\begin{figure}[h!]
\caption{Excess values (log scale), 'AL-BV' denotes the approach using the kernel smoothing estimator.  'AL-BV-DR (KR)' denotes the approach based on a doubly robust estimator using kernel regression for $Q^m(x,a)$ and 'AL-BV-DR (LR)' denotes the approach based on a doubly robust estimator using linear regression for $Q^m(x,a)$.}
\label{fig_dr}
\begin{center}
\includegraphics[width=2.1in,height=2.1in]{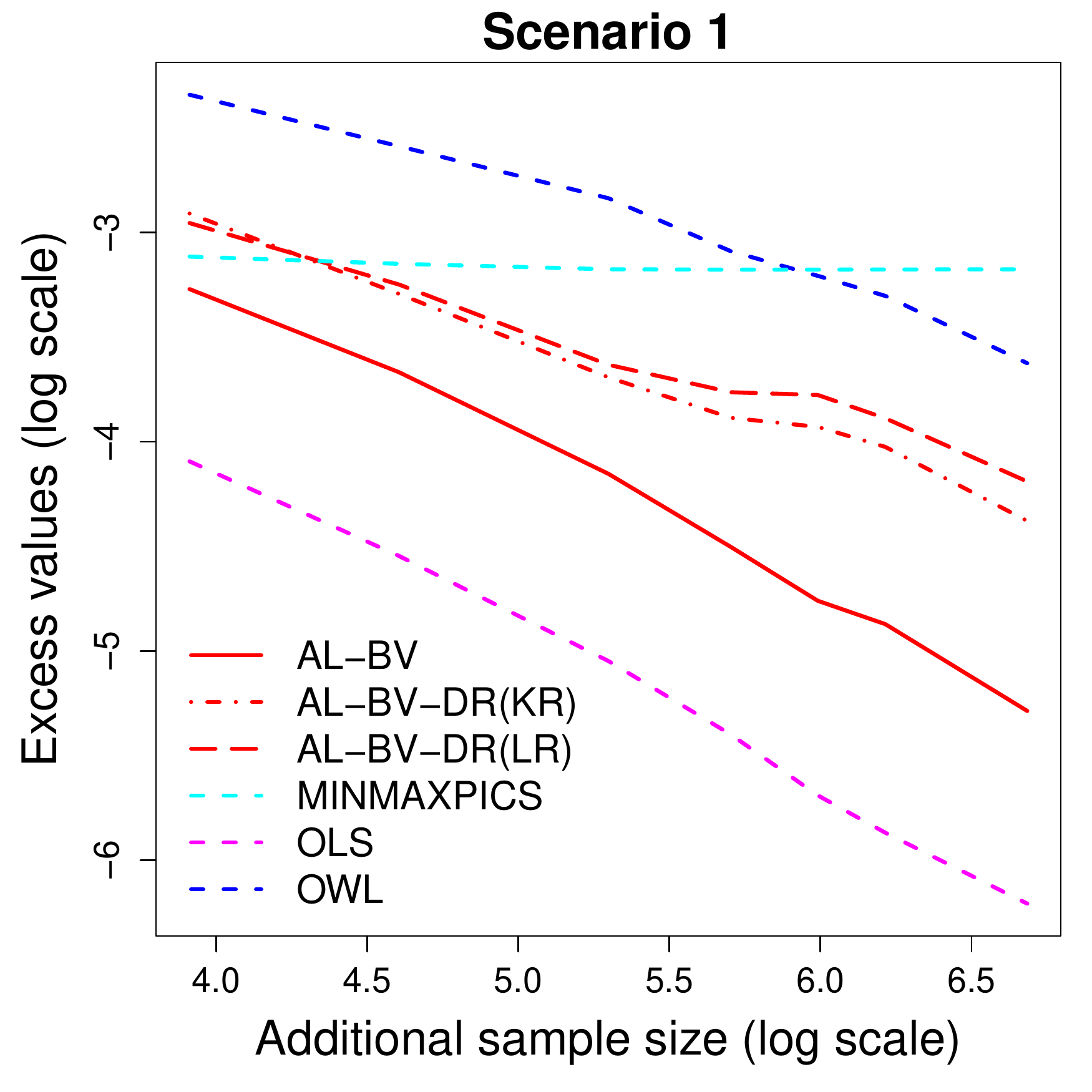}
\includegraphics[width=2.1in,height=2.1in]{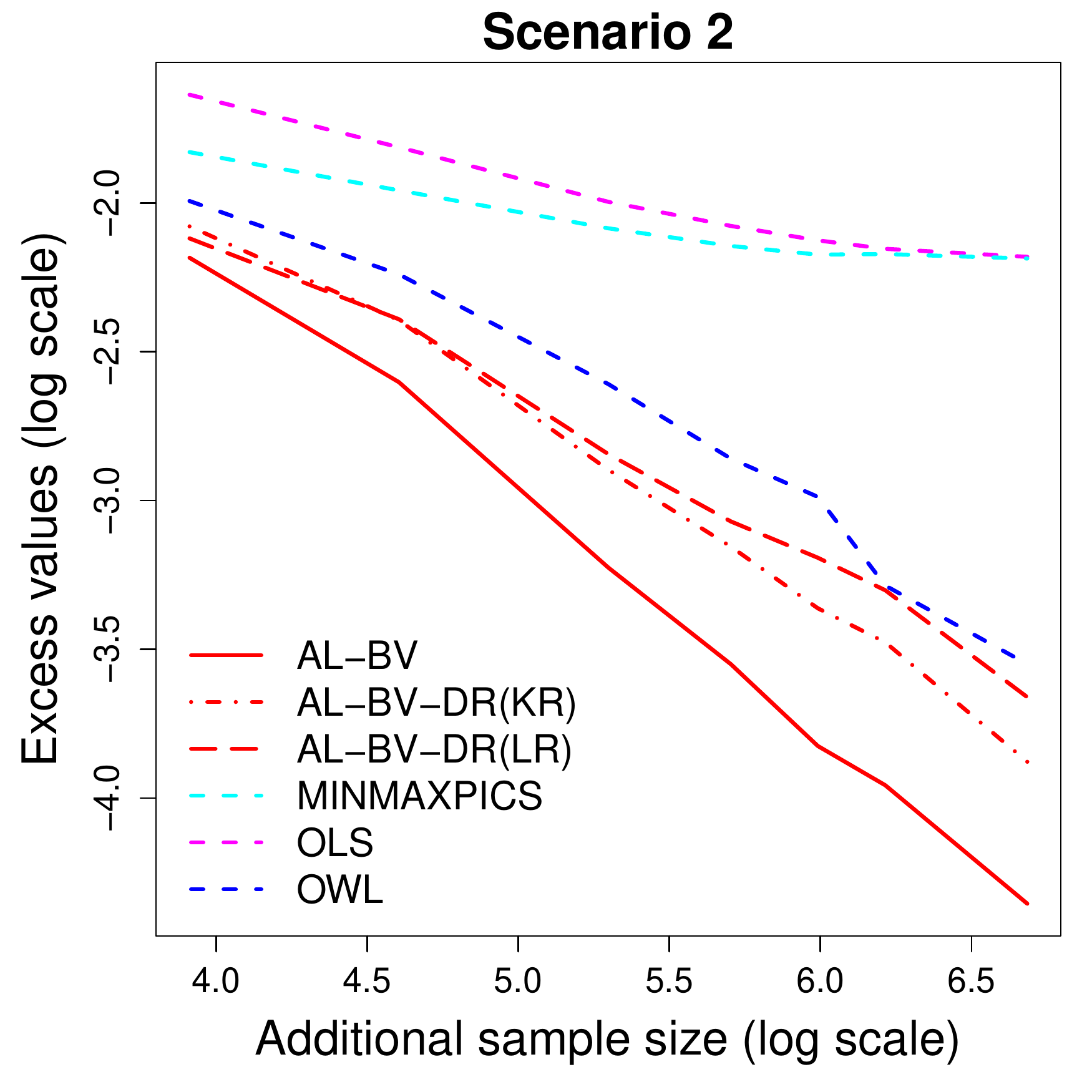}
\includegraphics[width=2.1in,height=2.1in]{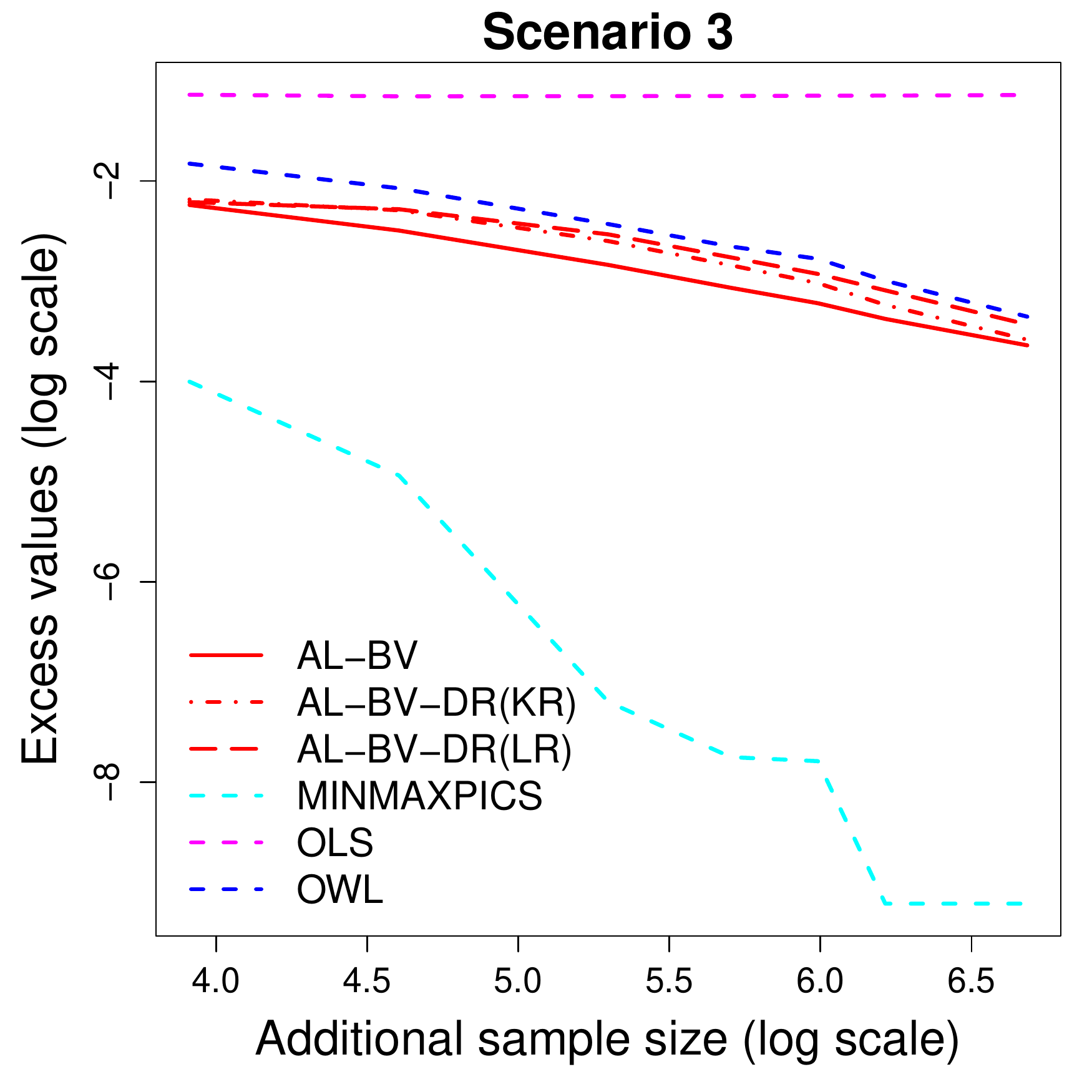}
\includegraphics[width=2.1in,height=2.1in]{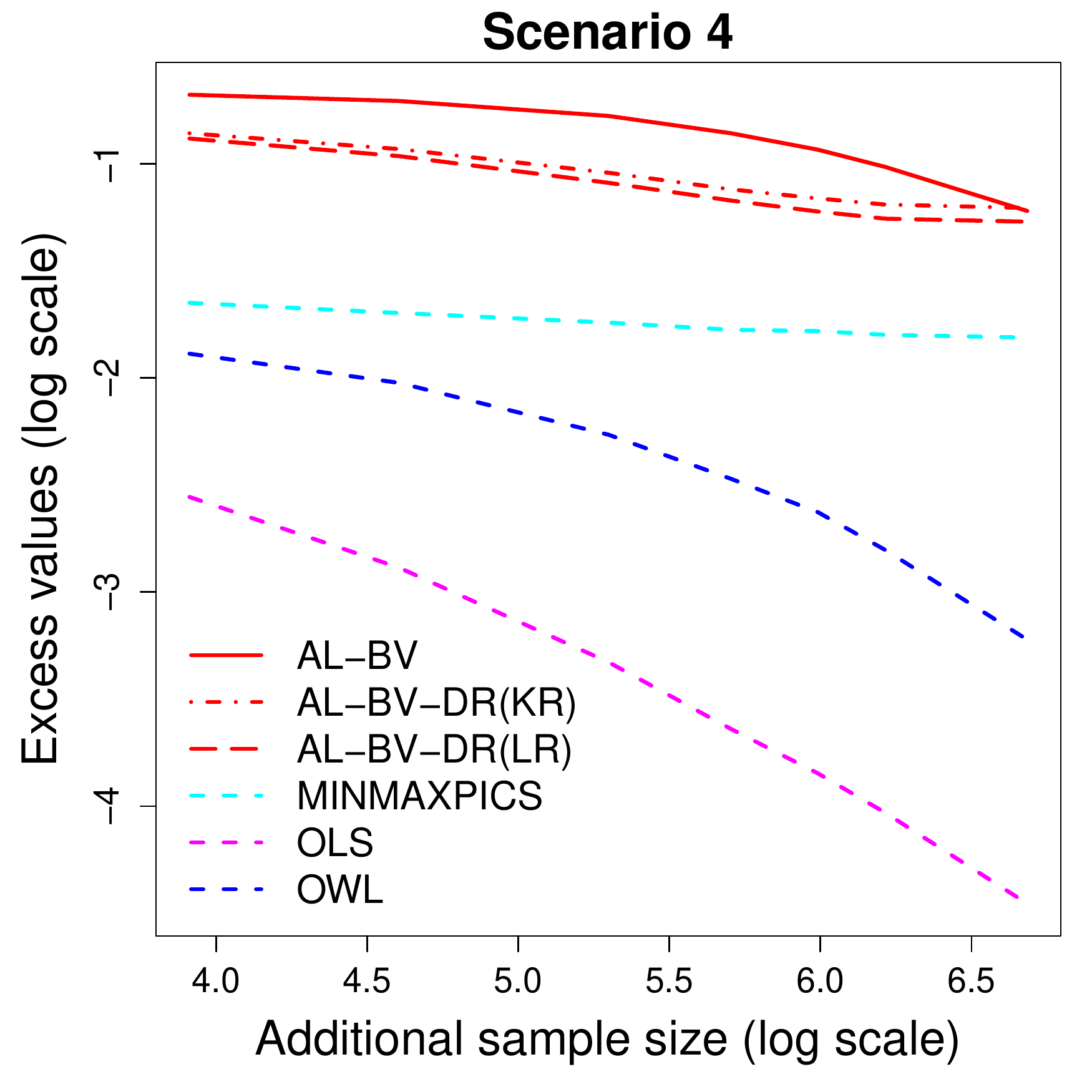}
\includegraphics[width=2.1in,height=2.1in]{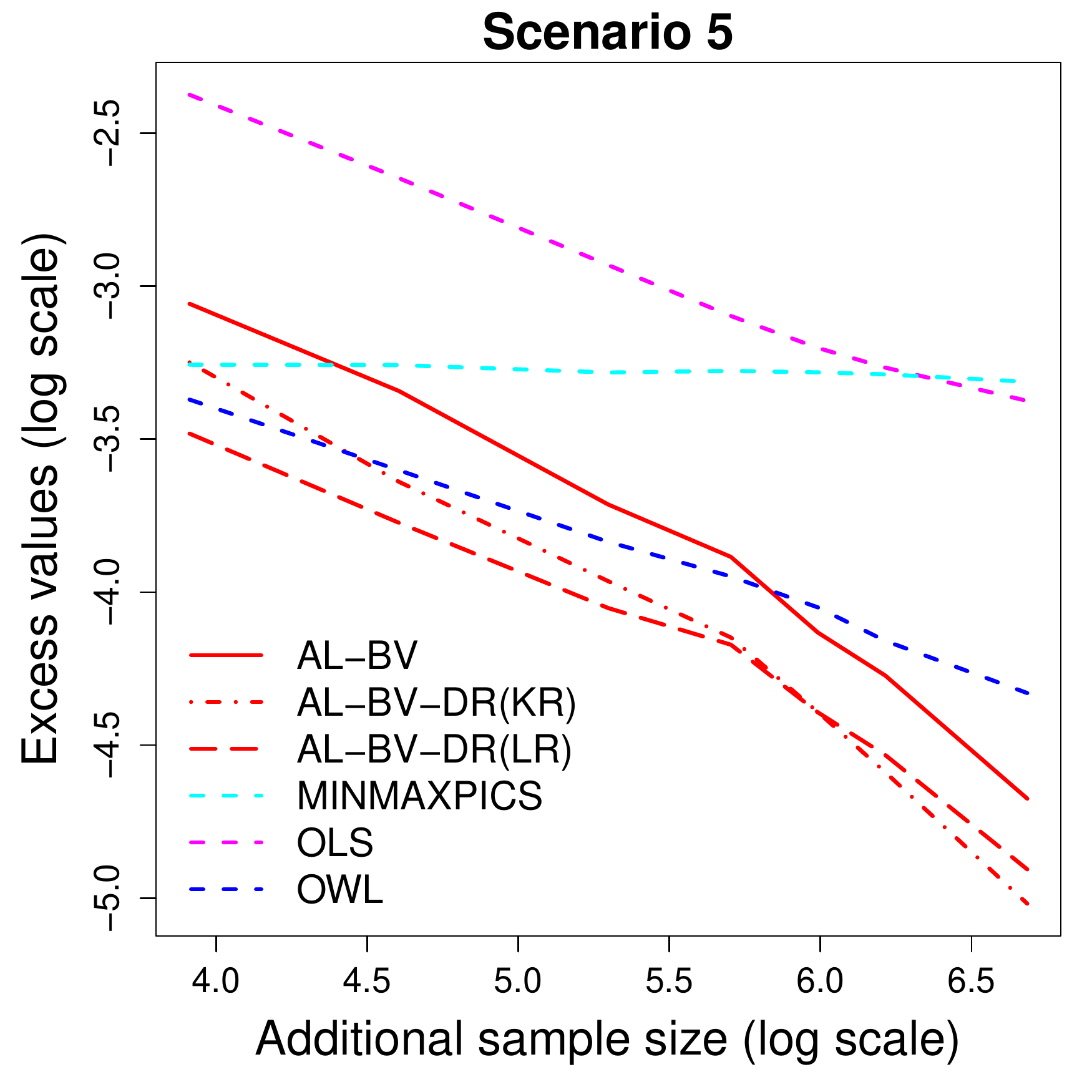}
\includegraphics[width=2.1in,height=2.1in]{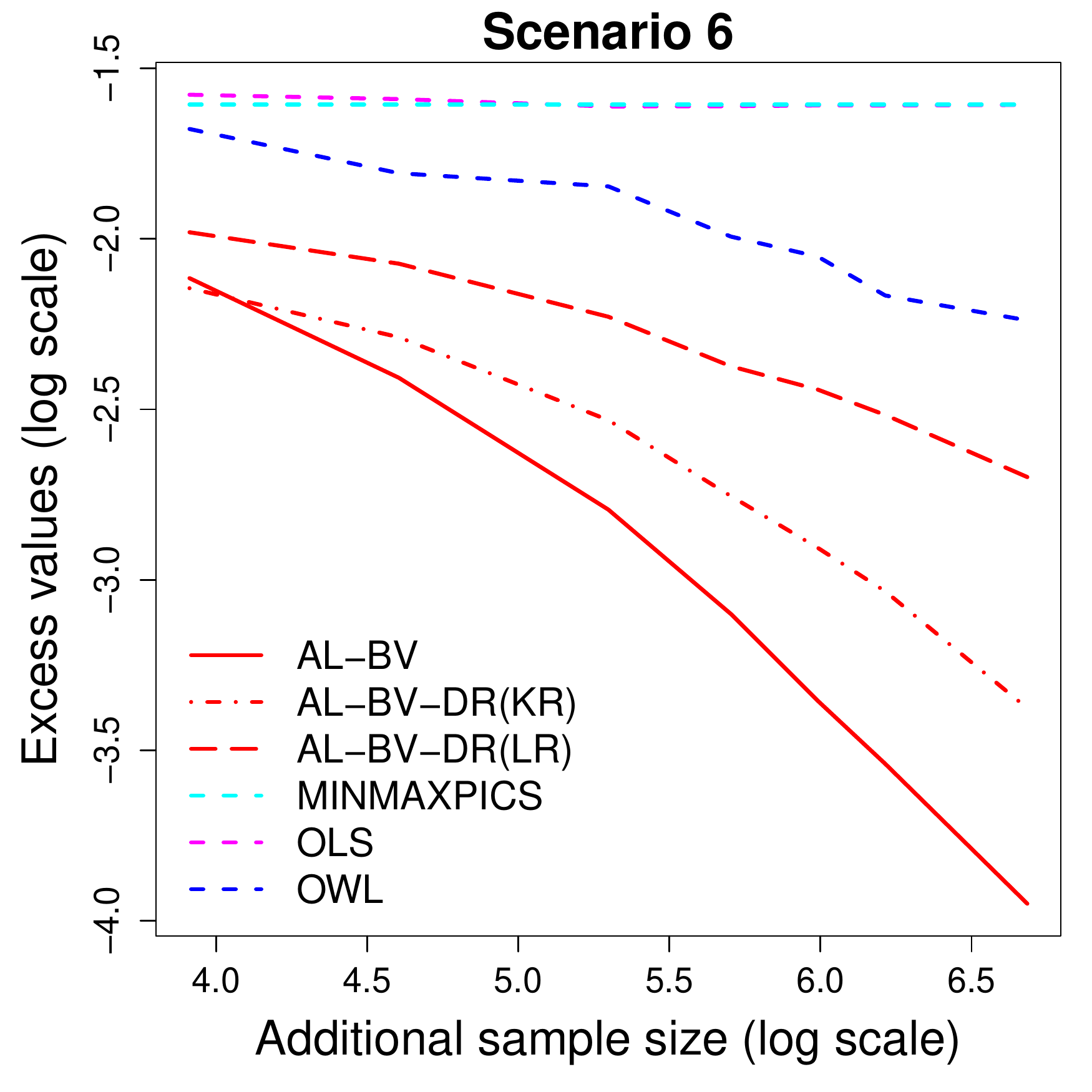}
\end{center}
\end{figure}

The individualized treatment rule can also be constructed via  a doubly robust augmented inverse probability weighted estimator. Here we examine the performance of such an estimator, which is used for estimation and participant inclusion in the active clinical trial.  A doubly robust representation for the value function $V(D)$ is
\begin{equation*}
V^{AIPWE,m}(D) \triangleq E\left[
\frac{RI\left\lbrace
  A=D(X)
\right\rbrace}{{\pi}^{m}(A;X)} -
\frac{I\left\lbrace A=D(X)\right\rbrace -  {\pi}^{m}\{D(X);X\}}
{{\pi}^{m}\{D(X);X\}}{Q}^{m}\{X, D(X)\}
\right],
\end{equation*}
where $\pi^{m}(a; x)$ denotes a working model for $P(A=a|X=x)$ and  $Q^{m}(x,a)$ denotes a working model for $E(R|X=x,A=a)$. It can be shown that $V^{AIPWE,m}(D) =V(D)$ when either $\pi^{m}(a; x)$  or $Q^{m}(x,a)$ is correctly specified \citep{Zhang:RobustOTR2012}. Denote
$$W_a^m =I(A=a)\frac{R-Q^m(X,a)}{\pi^m(a;X)}+Q^m(X,a).$$ The optimal treatment rule maximizing $V^{AIPWE,m}(D)$ is
$$
 D^*(x) = \sign\{E(W_1^m|X=x)-E(W_{-1}^m|X=x)\}.
$$
Within the framework of active clinical trial, patients for whom the difference between $E(W_a^m|X=x), a=\pm 1$, is small are more likely to get enrolled, since they are closer to the decision boundary. In the implementation, $\pi^m(a;x)=1/2$ is known. We can plug in the estimated $E(R|X=x,A=a)$ to estimate $E(W_a^m|X=x)$. Subsequently, the estimated  $E(W_1^m|X=x)-E(W_{-1}^m|X=x)$ can be used to select enrolled patients. We examine the performance of  active clinical trial  based on a doubly robust estimator, where $E(R|X=x, A=a)$ are estimated using either kernel regression method (`AL-BV-DR (KR)') or a simple linear regression with $(X, A, XA)$ as predictors (`AL-BV-DR (LR)'). Results for Scenarios 1-6 are shown in Figure  \ref{fig_dr}. These results are relatively insensitive to the choices of working models for $Q^m(x,a)$. However, the performances of the doubly robust estimators vary across different situations. This might be due to a larger variability from using the doubly robust estimators with more estimation required. It will be of interest to further investigate the properties of active clinical trial via doubly robust estimators.

\section*{S.2 Additional theoretical results}

\bl \bf Proof of Lemma A.1 \el

We prove Lemma A.1 here. Below is the usual ``bias-variance'' decomposition
\[
\|\widehat\eta_j(x;h)-\eta_j(x)\|_\infty \leq \|\widehat\eta_j(x;h)-\mb E\widehat\eta_j(x;h)\|_\infty+\|\eta_j(x)-\mb E\widehat\eta_j(x;h)\|_\infty.
\]
We start with a bound on the bias term. Note that for any $x\in S\cap\supp(\Pi)$,
$\mb E\widehat\eta_j(x;h)=\frac{1}{Q_h(x|S)}\int K\l(\frac{x-y}{h}\r)\eta_j(y)d\Pi_S(y)$, hence
\begin{align}
\label{eq:a10}
&\nonumber
\l|\widehat\eta_j(x;h)-\mb E\widehat\eta_j(x;h)\r|=\l|\frac{1}{Q_h(x|S)}\int (\eta(x)-\eta(y))K\l(\frac{x-y}{h}\r)d\Pi(y) \r|\leq \\
&
\frac{L}{Q_h(x|S)}\int \|x-y\|_2 K\l(\frac{x-y}{h}\r)d\Pi_S(y)=L\frac{Q_{h,1}(x|S)}{Q_h(x|S)}\leq L\frac{c_4}{c_3}h,
\end{align}
where $c_3, c_4$ are defined in Section A.3.1. Recall that $Q_{h,1}(x|S)$ is defined in (A.3) with
\begin{eqnarray}
\mb Q_{h,1}(x|S):=\int\limits_{\mb R^p} \|x-y\|_2  K_h\l({x-y}\r)d\Pi_S(y)\nonumber.
\end{eqnarray}


Next, we will bound the stochastic term. Given $\eps>0$, let $\{x^{(k)}\}_{k=1}^{N(\eps)}$ be the minimal $\eps$-net on $S\cap \supp(\Pi)$, so that for any $y\in S\cap\supp(\Pi)$ there exists $1\leq k(y)\leq N(\eps)$ such that
$\|y-x^{(k(y))}\|_2\leq \eps$.
Set $\eps:=h^{d+1}$. Then
\begin{align}
\label{eq:a20}
&\sup\limits_{y\in S\cap \supp(\Pi)}|\widehat\eta_j(y;h)-\eta_j(y)|=\\
& \nonumber
\sup_{y\in S\cap\supp(\Pi)}|\widehat\eta_j(y;h)-\widehat\eta_j(x^{(k(y))};h)+\widehat\eta_j(x^{(k(y))};h)-\eta_j(x^{(k(y))})+
\eta_j(x^{(k(y))})-\eta_j(y)|\leq \\
& \nonumber
\max_{1\leq i\leq N(\eps)}|\widehat\eta_j(x^{(i)};h)-\eta_j(x^{(i)})|+\sup_{\|x-y\|_2\leq \eps}|\eta_j(x)-\eta_j(y)|+
\sup_{\|x-y\|_2\leq \eps}|\widehat\eta_j(x;h)-\widehat\eta_j(y;h)|.
\end{align}
Lipschitz condition on $\eta_j$ implies that
\[
\sup_{\|x-y\|_2\leq h^{d+1}}|\eta_j(x)-\eta_j(y)|\leq L h^{d+1}
\]
and, since $|R|\leq M$ almost surely and $K$ has Lipschitz constant $L_K$,
\[
\sup_{\|x-y\|_2\leq h^{d+1}}|\widehat\eta_j(x;h)-\widehat\eta_j(y;h)|\leq \frac{ML_K \Pi(S)}{c_3 h^d}\eps=\frac{M L_K\Pi(S)}{c_3}h.
\]
Fix $1\leq i\leq N(\eps)$. We will apply the Bernstein's inequality (e.g., Lemma 2.2.9 in \citet{van1996weak}) to estimate
$|\widehat\eta_j(x_i;h)-\mb E\widehat\eta_j(x_i;h)|$.
Our assumptions imply that for all $1\leq i\leq N$,
\[
\l|R^{(i)} I\{A^{(i)}=j\}\frac{K_h\l(x-X^{(i)}\r)}{Q_h(x|S)P(A^{(i)}=j)}\r|\leq \frac{2M\|K\|_\infty \Pi(S)}{c_3 h^d}
\]
almost surely, hence
$$\l|R^{(i)} I\{A^{(i)}=j\}\frac{K_h\l({x-X^{(i)}}\r)}{Q_h(x|S)P(A^{(i)}=j)}-\mb ER^{(i)} I\{A^{(i)}=j\}\frac{K_h\l({x-X^{(i)}}\r)}{Q_h(x|S)P(A^{(i)}=j)}\r|\leq \frac{4M\|K\|_\infty \Pi(S)}{c_3 h^d}$$ almost surely for all $x\in S\cap \supp(\Pi)$.
Moreover, (A.4) implies that
\[
{\rm Var} \left(\widehat\eta_j(x;h)\right)\leq \frac{2M^2\|K\|_\infty}{n} \int \frac{K\l(\frac{x-y}{h}\r)}{Q^2_h(x|S)}d\Pi_S(y)
\leq \frac{2M^2\|K\|_\infty \Pi(S)}{c_3 n h^d}.
\]
Bernstein's inequality implies that for all $t>0$,
\begin{align*}
&
|\widehat\eta_j(x^{(i)};h)-\mb E\widehat\eta_j(x^{(i)};h)|\leq 2\max\l(M\sqrt{\frac{2\|K\|_\infty \Pi(S) t}{c_3 nh^d}}, 4\frac{M\|K\|_\infty t}{c_3 n h^d}\r)
\end{align*}
with probability $\geq 1-2e^{-t}$.
Combined with the union over all $1\leq i\leq N(\eps)$ and noting that $N(\eps)\leq \frac{C\Pi(S)}{\eps^d}$, we get that
\begin{align}
\nonumber
\max_{1\leq i\leq N(\eps)}&|\widehat\eta_j(x^{(i)};h)-\mb E\widehat \eta_j(x^{(i)};h)|\leq \\
&\nonumber
C\max\l(M\sqrt{\frac{2\|K\|_\infty \Pi(S) (t+d^2\log(1/h))}{c_3 nh^d}}, 2\sqrt{2}\frac{M\|K\|_\infty \Pi(S)(t+d^2\log(1/h))}{c_3 n h^d}\r)
\end{align}
with probability $\geq 1-2e^{-t}$. Combined with (\ref{eq:a10}) and (\ref{eq:a20}), this implies the result.
\bibliographystyle{jasa}
\bibliography{active}

 \end{document}